\documentclass[useAMS,usenatbib]{mn2e}
\usepackage{deluxetable}
\usepackage{epsf}
\usepackage{amsmath}

\newcommand{\kmsmpc}{\kms\;{\rm Mpc}^{-1}}
\newcommand{\lya}{Ly$\alpha$\ }

\newcommand{\hmpc}{h^{-1}{\rm Mpc}}

\newcommand{\kms}{\;{\rm km}\,{\rm s}^{-1}}
\newcommand{\msun}{M_\odot}
\newcommand{\sfr}{\dot{M_*}}
\newcommand{\Av}{A_V}
\newcommand{\mstar}{M_*}
\newcommand{\zstar}{Z_*}
\newcommand{\zsun}{Z_\odot}
\newcommand{\smyr} {\msun \mbox{ yr}^{-1}}

\newcommand{\parname}[1]{\noindent\emph{\underline{#1:} }}
\newcommand{\spoc}{{\sc Spoc}}
\newcommand{\thtsfr}{t_H/t_{SFR}}
\newcommand\ion[2]{#1$\;${\scshape{#2}}}

\title[Constraining High-z Galaxies Using Simulations]{Constraints on
Physical Properties of $z\sim 6$ Galaxies Using Cosmological Hydrodynamic
Simulations}

\begin{document}

\author[Finlator, Dav\'e, Oppenheimer]{Kristian Finlator, Romeel Dav\'e, Benjamin D. Oppenheimer\\
Department of Astronomy, University of Arizona, Tucson, AZ 85721}

\maketitle

 \begin{abstract}
We conduct a detailed comparison of broad-band spectral energy
distributions of six $z\ga 5.5$ galaxies against galaxies drawn from
cosmological hydrodynamic simulations.  We employ a new tool called
\spoc, which constrains the physical properties of observed galaxies
through a Bayesian likelihood comparison with model galaxies.  We first
show that \spoc~self-consistently recovers the physical properties of
a test sample of high-redshift galaxies drawn from our simulations,
although dust extinction can yield systematic uncertainties at the
$\approx50\%$ level.  We then use \spoc~to test whether our simulations
can reproduce the observed photometry of six $z>5.5$ galaxies drawn from
the literature.  We compare physical properties derived from simulated
star formation histories (SFHs) versus assuming simple models such as
constant, exponentially-decaying, and constantly rising.  For five
objects, our simulated galaxies match the observations at least as
well as simple SFH models, with similar favored values obtained for the
intrinsic physical parameters such as stellar mass and star formation
rate, but with substantially smaller uncertainties.  Our results are
broadly insensitive to simulation choices for galactic outflows and
dust reddening.  Hence the existence of early galaxies as observed is
broadly consistent with current hierarchical structure formation models.
However, one of the six objects has photometry that is best fit by a
bursty SFH unlike anything produced in our simulations, driven primarily
by a high $K$-band flux.  These findings illustrate how \spoc~provides
a robust tool for optimally utilizing hydrodynamic simulations (or any
model that predicts galaxy SFHs) to constrain the physical properties
of individual galaxies having only photometric data, as well as identify
objects that challenge current models.
\end{abstract}

 
\section{Introduction}

Over the last few years, observations of galaxies at $z\sim
6$ have opened up a new window into the reionization epoch,
turning it into the latest frontier both for observational and
theoretical studies of galaxy formation.  Planned~\citep{gon05}
and existing wide-area narrowband searches for $z\ga 5.5$
objects such as the Subaru Deep Field~\citep{aji06,shi06}, the
Large Area Lyman-Alpha Survey~\citep{rho01,mal04}, the Chandra
Deep Field-South~\citep{wan05,mal05}, and the Hubble Ultra Deep
Field~\citep{mal05} are now combining with Lyman-alpha dropout
searches~\citep{dic04,bou04a,bou04b,mob05,bou06,eyl06,lab06}, targeted
searches near lensing caustics in galaxy clusters~\citep{kne04, hu02}
and occasionally serendipity~\citep{ste05} to uncover star-forming
galaxies from the reionization epoch in significant numbers \citep[see]
[for a listing of spectroscopically-confirmed $z>5$ galaxies]{ber06}.

A question immediately raised by this new stream of observations is,
what are the physical properties of these early galaxies?  Optimally, one
would determine properties such as the stellar mass, star formation rate,
and metallicity directly from high-quality spectra, but at present this is
infeasible for such faint systems.  Hence properties must be inferred from
photometry alone, occasionally augmented by emission line information.
This requires making some poorly constrained choices for the intrinsic
galaxy properties.  A commonly applied method known as Spectral Energy
Distribution (SED) fitting involves generating an ensemble of population
synthesis models under a range of assumptions for the intrinsic nature of
the object, and then finding the set of assumptions that best reproduces
a given galaxy's observed photometry~\citep[e.g.][]{ben00,kau03}.
The physical properties that yield the lowest $\chi^2$ model are then
forwarded as the most probable values, sometimes with little attention to
statistical uniqueness or robustness~\citep[see][for a nice exploration
of such issues]{sch05}.

Amongst the various assumptions used in SED fitting, the one that is often
least well specified and produces the widest range in final answers is
the galaxy's star formation history (SFH).  With no prior information,
common practice is to use simple SFHs with one free parameter such as
constant, single-burst, or exponentially-decaying, which in aggregate
are assumed to span the range of possible SFHs for a given galaxy.
Indeed, in most cases all one-parameter SFHs yield plausible results,
though the parameters obtained and quality of fits in each case can
vary significantly.  If it were possible to narrow the allowed range of
SFHs through independent considerations, physical parameters could 
in principle be more precisely determined.

One approach for constraining SFHs a priori is to incorporate information
from currently favored hierarchical structure formation models.  As we
will discuss in this paper, hydrodynamic simulations tend to produce a
relatively narrow range of star formation histories for early galaxies.
Their galaxies' SFHs tend to follow a generic {\it form} at these early
epochs, best characterized as a constantly-rising SFH.  This form
is broadly independent of cosmology, feedback assumptions, or other
ancillary factors, and is furthermore distinct from any one-parameter
models commonly used today.  A primary aim of this paper is to test
whether this relatively generic SFH form is consistent with observations,
and if so, what the implication are for the physical properties of
high-redshift galaxies.

Despite impressive recent successes in understanding cosmology and
large-scale structure in our Universe~\citep[e.g.][]{spe06,spr06},
many uncertainties remain in our understanding of galaxy formation.
Several recent papers have tested models of high-$z$ galaxy formation by
comparing them to observed bulk properties such as luminosity functions
at rest-frame UV and \lya wavelengths.  These comparisons have shown
that such models are broadly successful at reproducing observations,
under reasonable assumptions for poorly constrained parameters such as
dust extinction~\citep{som01,idz04,nig05,fin06,dav06a}.  While this
broad success is encouraging, it is subject to some ambiguousness in
interpretation, because the properties of individual galaxies are not
being compared in detail.  One could envision situations in which a model
reproduces an ensemble property of galaxies but not the detailed spectra
of individual objects.  As an example, it was forwarded by \citet{kol99}
that Lyman break galaxies at $z\sim 3$ are actually merger-driven
starbursts, in contrast to many other models predicting them to be large
quiescent objects.  Despite quite different SFHs, both models reproduced
many of the same bulk properties such as number densities and clustering
statistics.  For $z\ga 6$ galaxies where statistics are currently poor,
such degeneracies can hamper interpretations of bulk comparisons of
observations to models.

A complementary set of constraints on galaxy formation models may be
obtained by comparing models to the individual spectra of observed
galaxies.  In practice, for high-$z$ systems, photometry over a
reasonably wide set of bands must substitute for detailed spectra.
Such comparisons of models to data would move towards more precise
and statistically robust analyses that do not rely on having a large
ensemble of objects.  This last aspect is critical, because the very
earliest observed objects that may provide the greatest constraints on
models will in practice always be few in number and detected only at
the limits of current technology.

In short, what is desireable would be a tool to compare models and
observations of high-redshift galaxies that (1) employs reasonably
generic predictions of current galaxy formation models; (2) provides
a quantitative and robust statistical assessment of how well such
models reproduce observations; (3) yields information on the physical
properties of galaxies under various assumptions; (4) obtains such
information based solely on observed photometry; and (5) does all
this on a galaxy-by-galaxy basis rather than relying on having a large
statistical sample of observed galaxies.

In this paper we introduce such a tool, called \spoc~(Simulated
Photometry-derived Observational Constraints).  \spoc\ takes as its input
the photometry (with errors) of a single observed galaxy along with an
ensemble of model spectra drawn either from simulations or generated using
one-parameter SFHs.  The output is probability distributions of physical
parameters derived using a Bayesian formalism, along with goodness-of-fit
measures for any given model.  The probability distributions give
quantitative constraints on the galaxy's physical properties, while the
goodness-of-fit can be used to discriminate between models and determine
whether a given model (be it simulated or one-parameter SFHs) is able to
provide an acceptable fit to that galaxy's photometry.

After introducing and testing \spoc, we apply it to a sample of six
$z>5.5$ galaxies from the literature that have published near-infrared
photometry.  We show that in five of six cases, the simulated galaxies
fit observations at least as well as one-parameter SFHs.  Since there is
no guarantee that simulations produce galaxy SFHs that actually occur in
nature, the fact that good fits are possible shows that the existence
of the majority of observed $z\ga 5.5$ galaxies is straightforwardly
accommodated in current galaxy formation models.  However, in one case,
we find that simulated galaxies provides a much poorer fit than can be
obtained with one-parameter SFHs, as burstier SFHs provide a much better
fit than can be obtained from any simulated galaxies.  At face value,
this implies that our simulations cannot yet accommodate the full range
of observed galaxies, and that some physical process may be missing,
although we will explore alternate interpretations.  For each galaxy
we also present the best-fit physical parameters, with uncertainties,
obtained using each model SFH.  The simulations provide significantly
tighter constraints than the full range of one-parameter SFHs, as expected
based on their relatively small range of SFHs produced.  These values
can therefore be regarded as predictions of our simulations that may be
tested against future observations.

\S~\ref{sec:spoc} introduces \spoc, detailing our Bayesian formalism
and discussing systematic uncertainties.  \S~\ref{sec:models} presents
the simulations and the one-parameter models that will be used as the
template library for \spoc.  \S~\ref{sec:performance} discusses what
drives the inferred physical properties in the context of our simulations,
and shows that~\spoc~accurately recovers the physical properties of
simulated galaxies.  \S~\ref{sec:kesr} explores the best-fit parameters
of one observed reionization-epoch galaxy in detail, and compares with
results from traditional one-parameter SFH models.  \S~\ref{sec:otherGals}
repeats the previous comparison for a larger set of observed galaxies,
highlighting the variety of interesting results that \spoc\ obtains.
Finally, in \S~\ref{sec:summary} we present our conclusions.

\section{Methodology of \spoc} \label{sec:spoc}

\subsection{SED Fitting}\label{sec:sedfitting}

Pedagogical explanations of SED fitting techniques have been presented
elsewhere~\citep{ben00,kau03}, so we refer the reader there for more
detailed discussion of those aspects.  Here we provide some basic
insights and notes.

Clearly, the amount of physical information that can be inferred
from available data depends on the quantity and quality
of the data.  For some high-$z$ galaxies, only narrow-band
photometry and rest-frame ultraviolet (UV) spectroscopy are
available~\citep[e.g.,][]{cub03,kod03,rho03,kur04,rho04b,ste05,wes05}.
For others, an emission-line measurement and 1--3 rest-UV broad bands
are available~\citep[e.g.,][]{nag04,sta04a,sta04b,nag05,sti05,hu04}.
Studies employing the Lyman dropout technique in the optical
must further contend with the possible presence of low-redshift
interlopers~\citep{dic04,bou04a,bou04b} and large uncertainties from
dust extinction.  Nevertheless, some interesting constraints can be
placed on the underlying physical properties of the sources from solely
rest-UV data~\citep{dro05,gwy05}.

With the addition of rest-frame optical data, e.g. from \emph{Spitzer's}
Infrared Array Camera (IRAC), it becomes possible to obtain simultaneous
constraints for the stellar mass, star formation rate (SFR), dust
extinction, and redshift using spectral energy distribution (SED) fitting
techniques~\citep{ega05,cha05,eyl05,mob05,yan05,sch05,dun06,lab06}.
The uncertainties inherent in such analyses primarily stem from a poor
constraint on the age of the galaxy's stellar population, because the
relationship between age and the strength of the telltale Balmer break
depends on the form of the assumed SFH (\citealt{pap01,sha05}; 
Figure~\ref{fig:degeneracy}).
This age uncertainty propagates via a host of degeneracies into increased
uncertainties in the inferred stellar mass, SFR, metallicity, and dust
extinction, if no priors are assumed on these quantities.  Additional
uncertainties arise from the unknown form of the appropriate template
SED~\citep[e.g.][]{sch05} and the treatment of stellar evolution assumed
by the chosen population synthesis models~\citep[see e.g.][]{mar06}.
Still, SED fitting offers the most promising approach for determining
the physical properties of individual high-$z$ galaxies.

Given this, how can one employ simulations to improve constraints on
SED fitting?  One can view a numerical simulation as producing a Monte
Carlo sampling of parameter space such that the frequency with which
a given set of physical parameters ought to occur is proportional to
the number of galaxies in the simulation that are characterized by that
set of parameters.  In essence, numerically-simulated galaxies provide
``implicit priors" for SED fitting, i.e. solutions that are \emph{a
priori} weighted more heavily because they occur more frequently.

The underlying assumption is that simulated galaxy SFHs represent
those occuring in nature.  This is by no means guaranteed, and indeed
whether \spoc\ provides an acceptable fit to a given galaxy constitutes
a stringent test of the simulation, because a galaxy's spectrum encodes
information about its full SFH.  This is the manner in which \spoc\
can provide a test of galaxy formation models based on individual 
systems.

\subsection{The \spoc\ Equation} \label{sec:stats}

We now summarize the Bayesian statistical method employed in \spoc.  
Our goal is to constrain the stellar mass, SFR, mean stellar metallicity,
age, dust extinction, and redshift ($\mstar$, $\sfr$, $\zstar$, $t$,
$\Av$, and $z$, respectively) based on available measurements $D$.
According to Bayes' Theorem, the probability $p$ that the measurements
$D$ correspond to a galaxy with the intrinsic physical parameters
$\hat\phi\equiv(\hat{\mstar}, \hat{\sfr}, \hat{\zstar}, \hat{t},
\hat{\Av}, \hat{z})$ (where a hat indicates a particular value of a
parameter) is given by
\begin{equation} \label{bayes1}
p(\hat{\phi}\mid D) \propto p(\hat{\phi}) p(D \mid \hat{\phi}).
\end{equation}
The prior $p(\hat{\phi})$ indicates the relative \emph{a priori}
probability that a randomly selected galaxy has this particular
combination of parameters, and the likelihood $p(D \mid \hat{\phi})$
indicates the probability of obtaining the measurements $D$ for a
galaxy characterized by the parameters $\hat{\phi}$; for a given
model galaxy and data set $D$ this is assumed to be proportional to
$e^{-\chi^2/2}$.  Any information regarding the expected distributions
of physical properties of the observable galaxies (such as the stellar
mass function) or relationships between these properties (such as a
mass-metallicity relation) can be taken into account via a contribution
to the prior, and will generally give rise to more precise---and possibly
more accurate---constraints.

In this work, we assume uniform priors on $z$ and $\Av$, and we do not
assume any dependence between $\Av$ and the other intrinsic physical
properties.  We introduce an additional prior $p(\mbox{sim})$ to account
for any other priors.  For example, when matching observed galaxies
against model galaxies derived from the outputs of two cosmological
simulations that span different comoving volumes, $p(\mbox{sim})$
represents the ratio of the simulation volumes.  After several
applications of the product rule, we obtain
\begin{equation} \label{bayes2}
p(\hat{\phi}) \propto
p(\hat{\mstar},\hat{\sfr},\hat{\zstar},\hat{t} \mid \hat{z})
p(\mbox{sim})
p(D \mid \hat{\phi}).
\end{equation}
This is the fundamental equation that \spoc~evaluates.  Generically,
one would use Equation~\ref{bayes2} by beginning with a set of models
that uniformly samples the relevant parameter space and then guessing
the form of the prior $p(\hat{\mstar},\hat{\sfr},\hat{\zstar},\hat{t}
\mid \hat{z})$, which now encodes the assumed distribution of intrinsic
physical properties of galaxies as a function of redshift.  In the
high-redshift literature, where little is known about the intrinsic
physical properties of the galaxies, it is common to neglect priors
altogether (or, equivalently, to choose the model with the lowest
$\chi^2$) or even to introduce them accidentally by not sampling parameter
space uniformly.  The difference between this work and that of previous
authors is that we account for this prior implicitly by using numerically
simulated galaxies as the model set.

To see how this works, consider how one would use equation~\ref{bayes2}
in practice.  For simplicity, suppose that we wished to constrain a
galaxy's stellar mass and that the mass could only fall within one of
two ranges.  If we omitted priors and assumed that the models sample
stellar mass uniformly, then the probability that the galaxy's mass falls
within a given range would be given by $\sum_i A e^{-\chi_i^2/2}$, where
the sum is taken over all models whose mass lies within that range and
the normalization $A$ is chosen so that the sum taken over all models
in both ranges equals unity.  If we believed that galaxies with masses
in one range were, say, twice as common (and therefore \emph{a priori}
twice as likely to be the right answer) as galaxies with masses in the
other range, we could account for this via an explicit prior by changing
the sum to $\sum_i A P_i e^{-\chi_i^2/2}$ where $P_i = 2$ for models
in the more common range and $1$ for models in the less common range
(with $A$ of course rescaled).  It is clear that an equivalent method to
employing this explicit prior would be to generate twice as many models
in the more common range, resulting in twice the probability of selecting
one of these models.  Generalizing this idea, one can view simulated
galaxies as a Monte Carlo sampling of parameter space that naturally
produces more models with parameters that are more commonly found.
Hence by taking a set of simulated galaxies, generating a library by
resampling this set with parameters having uniform priors (namely, $A_V$
and $z$), and using that library to discretely sample the probability
distribution in the right-hand side of equation~\ref{bayes2}, one can
solve equation~\ref{bayes2} effectively incorporating the implicit
priors given by the simulated galaxies.  This is in essence the \spoc\
algorithm.

\subsection{Systematic Uncertainties in Using Simulated Galaxies} \label{sec:strengths}

A major difficulty with the \spoc\ approach is that there is no guarantee
that the simulation predicts the correct distribution of intrinsic
properties of galaxies; in Bayesian terms, the priors could be wrong.
On some level this is bound to be the case as we do not account for every
process that could \emph{in principle} affect galaxies at this epoch;
indeed, no model currently does.  However, our goal is to determine
whether our treatment is \emph{sufficient} to account for current
observations.  If not, then the failures indicate needed improvements to
the model.  If our treatment can account for current observations, then
the constraints that we derive may be regarded as physically-motivated
predictions, subject to verification when more constraining data become
available.

The two greatest uncertainties for the input physics present in our
current simulations are (1) numerical resolution---manifested either as
an inability to account for physical processes that occur on scales that
are too small or too rapid (e.g. merger-induced starburst) or as a lack of
numerical convergence---and (2) the prescription for superwind feedback.
In \S~\ref{sec:resolution} we use a simple convergence test to argue
that our results do not suffer from numerical resolution limitations.
As to our treatment for outflows, we can estimate the extent of any
resulting systematics by comparing results from our three different
outflow simulations.  While this does not span the full range of possible
feedback mechanisms, the fact that (as we show in \S~\ref{sec:windDust})
most of the best-fit parameters are insensitive to the choice of wind
prescription suggests that outflows do not noticeably alter typical SFHs
at a given stellar mass.

On the other hand, if there are significant physical processes affecting
galaxy SEDs that are not accounted for by our simulation or population
synthesis models, then our simulated galaxies may fail to reproduce the
observed spectra, or they may mistakenly model nonstellar contributions
to the observed SED as starlight.  Among the possibilities here are
active galactic nuclei (AGN), incorrectly modeled thermally pulsating
asymptotic giant branch (TP-AGB) stars~\citep{mar06}, emission lines,
and an inappropriate treatment of dust or IGM absorption.  We will argue
in \S\ref{sec:otherGals} that significant AGN contamination is unlikely
for the high-redshift objects we will consider here.  The contribution
of TP-AGB stars is also unlikely to be important partly because we do
not model measurements from bands redder than $I$ in the rest-frame, and
partly because at $z\sim6$ less than half of the existing stellar mass
is more than 200 Myr old (Table~\ref{table:many_objs}).  Emission lines
and incorrectly-modelled IGM absorption could in principle affect our
results at the 10\% level~\citep{sch05,ega05}.  These effects are expected
to be similar for the various SFHs investigated because we use the same
population synthesis models to model the stellar continuum in each case.
Regarding dust, we have found, in agreement with~\citet{sch05}, that
our results are relatively insensitive to the form of the dust law that
we consider (see \S~\ref{sec:consistency}).  Thus, for the preliminary
study in this paper we ignore all of these effects.

Another possible problem is that galaxy classes that are rare in
reality are likely to be rare in the simulations.  Accordingly, if the
comoving volume from which the catalog of simulated comparison galaxies
is drawn is sufficiently small that a simulated analogue to an observed
rare object is neither expected nor found, that object cannot directly
constrain the model.  For example, our simulations produce no galaxies
massive enough at $z>6$ to fit HUDF-JD2, the putative $6\times10^{11}
\msun$ object at $z\sim6.5$ reported by~\citet{mob05}.   Although this
particular object is likely to be at a lower redshift \citep{dun06},
it does illustrate limitations imposed by simulation volume, which
could also impact constraints on rare classes such as sub-millimeter
galaxies \citep[e.g.][]{sma04} (alternatively, if such objects are indeed
common at $z \geq6$ then they represent a challenge to our simulations).
In principle one could work around this issue by running larger-volume
simulations or by deriving the priors from the simulations and then
resampling parameter space by hand.  In lieu of these approaches,
the simulations utilized must have comoving volumes comparable to the
effective volume of the survey in which the object was found.

It may appear overly ambitious to attempt to constrain 6 (or more)
seemingly independent parameters for a galaxy for which fewer than 6
measurements are available.  However, cosmological simulations allow
us to do this because they generically predict that galaxies' intrinsic
physical parameters are manifestly \emph{not} independent; there are tight
predicted correlations between, for example, stellar mass on the one hand
and star formation rate and metallicity on the other~\citep{fin06,dav06a}.

In summary, using simulated galaxies to estimate physical properties
is only valid when the dominant emission mechanism is star formation,
and when other uncertainties can be carefully analyzed and shown to be
negligible.  For galaxies at high redshift, such as the ones we consider
in this paper, this is believed (but not guaranteed) to be true.  However,
in the general case these issues must be considered carefully.  In turn,
the goodness of fit enables constraints to be placed on simulations of
galaxy formation, and can highlight missing physics that may be required
in order to explain the observed properties of galaxies.

\section{Models} \label{sec:models}
\subsection{Simulations} \label{sec:simulations}

We draw our simulated galaxies from cosmological hydrodynamic
simulations run with Gadget-2~\citep{spr02}, including our improvements
as described in~\citet[][hereafter OD06]{opp06}.  This code uses an
entropy-conservative formulation of smoothed particle hydrodynamics
(SPH) along with a tree-particle-mesh algorithm for handling
gravity.  Heating is included via a spatially uniform photoionizing
background~\citep{haa01}, which is an acceptable approximation for
the galaxies that are observed at high redshift owing to the fact that
they form in highly overdense regions that undergo local reionization
at $z\gg 6$~\citep{dav06a}.  All gas particles are allowed to cool under
the assumption of ionization equilibrium, and metal-enriched particles
may additionally cool via metal lines.  Cool gas particles are allowed to
develop a multi-phase interstellar medium via a subresolution multi-phase
model that tracks condensation and evaporation following~\citet{mck77}.
Stars are formed from cool, dense gas using a recipe that reproduces the
\citet{ken98} relation; see \citet{spr03a} for details.  The metallicity
of star-forming gas particles grows in proportion to the SFR under the
instantaneous recycling approximation.  Stars inherit the metallicity
of the parent gas particle, and from then on cannot be further enriched.

Cosmological hydrodynamic simulations that do not include
kinetic feedback from star formation invariably overproduce
stars~\citep[e.g.,][OD06]{bal01, spr03a}.  Because superwinds can affect
the physical properties of the simulated galaxies~\citep[e.g.][]{dav06a},
we consider model galaxies from simulations with three different superwind
schemes: (1) a ``no wind" model that omits superwind feedback; (2) a
``constant wind" (cw) model in which all the particles entering into
superwinds are expelled at 484~km/s out of star forming regions and
a constant mass loading factor (i.e. the ratio of the rate of matter
expelled to the SFR) of 2 is assumed \citep[as in the runs of][]{spr03b};
and (3) the ``momentum-driven wind" (vzw) model of~OD06, in which the
imparted velocity is proportional to the local velocity dispersion
(computed from the potential) and the mass loading factor is inversely
proportional to the velocity dispersion \citep{mur05}, as inferred from
observations of local starbursts \citep{mar05,rup05}.  This selection
is meant to bracket plausible models in order to expose any related
systematic uncertainties; however, owing to the range of successes in
comparison with IGM metal-line observations obtained by~OD06 for the vzw
model, we focus on this model when the conclusions from the different
wind models are broadly similar.

All of our wind models were tested in simulations that assumed the
``old" WMAP-concordant cosmology \citep{spe03} having $\Omega=0.3$,
$\Lambda=0.7$, $H_0=70\kmsmpc$, $\sigma_8=0.9$, and $\Omega_b=0.04$.
Each of our simulations has $2\times 256^3$ particles, with parameters as
given in OD06.  We only employ the $16\hmpc$ and $32\hmpc$ simulations
from OD06, as the $8\hmpc$ runs did not have any galaxies large enough
to be observable at $z\ga 6$.  An additional set of simulations (the
``jvzw" model) were run using our preferred wind model with the 3rd-year
WMAP cosmology~\citep{spe06}, namely $\Omega=0.26$, $\Lambda=0.74$,
$H_0=71\kmsmpc$, $\sigma_8=0.75$, and $\Omega_b=0.044$.  Due to an error
in the initial conditions generation, the power spectrum index was set
to $n=1$ rather than the currently-favored $n=0.95$; however, this has
little impact on our results as we will show that they are insensitive
to such differences in cosmology.  There is a slight change in the
wind model for jvzw versus vzw, in that jvzw has a smaller mass loading
factor by a factor of two-thirds compared to vzw (in the terminology of
OD06, $\sigma_0=200$~km/s) in order to compensate for the lower collapse
fraction at high redshift in the new cosmology.  In addition to $16\hmpc$
and $32\hmpc$ box sizes, we also run a $64\hmpc$ box with the jvzw model
to sample the bright end of the mass function in order to better constrain
some observed galaxies that we will consider in \S\ref{sec:otherGals}.
We found that model galaxies from the jvzw simulations have bulk
properties that are similar to that from vzw.  For this paper we will
compute all luminosity distances assuming the new 3rd-year WMAP cosmology.

We identify galaxies using Spline Kernel Interpolative DENMAX
(see~\citealt{ker05} for a full description).  We only consider galaxies
with stellar masses exceeding 64 star particles, which represents
a converged sample in terms of both stellar mass and star formation
history~\citep{fin06}.  According to this criterion, our $16\hmpc$
simulation volumes resolve galaxies with stellar mass $\ga 1.2 \times
10^8 \msun$.

For this work, the most important output of the simulations is the set
of SFHs corresponding to the resolved galaxies in each simulation at the
various redshift outputs.  We obtain the rest-frame spectrum for each
star formation event in a given galaxy at the time of observation by
interpolating to the correct metallicity and age within the~\citet{bru03}
models, assuming a Chabrier IMF.  Summing these up, we obtain the galaxy's
intrinsic rest frame spectral energy distribution (SED).

We consider the following prescriptions for dust reddening: The
\citet{cal00} starburst dust screen, the~\citet{cf00} embedded star
formation law, the~\citet{gor03} Small Magellanic Cloud bar law,
and the~\citet{ccm89} Milky Way law.  We account for IGM absorption
bluewards of rest-frame \lya using the~\citet{mad95} prescription.
The~\citet{mad95} law may be less appropriate for $z\ga 6$ than at lower
redshifts because the universe is completing reionization at this epoch.
Indeed, \citet{sch05} found that they were able to improve the quality
of their fits to the SEDs of two reionization-epoch galaxies by simply
doubling the optical depth predicted by~\citet{mad95}.  However, they also
found that the best-fit derived parameters are relatively insensitive
to the IGM treatment.  Thus, for simplicity we retain the~\citet{mad95}
treatment without modification.

\subsection{One-Parameter Star Formation Histories} \label{sec:SFHs}

\begin{figure}
\setlength{\epsfxsize}{0.5\textwidth}
\centerline{\epsfbox{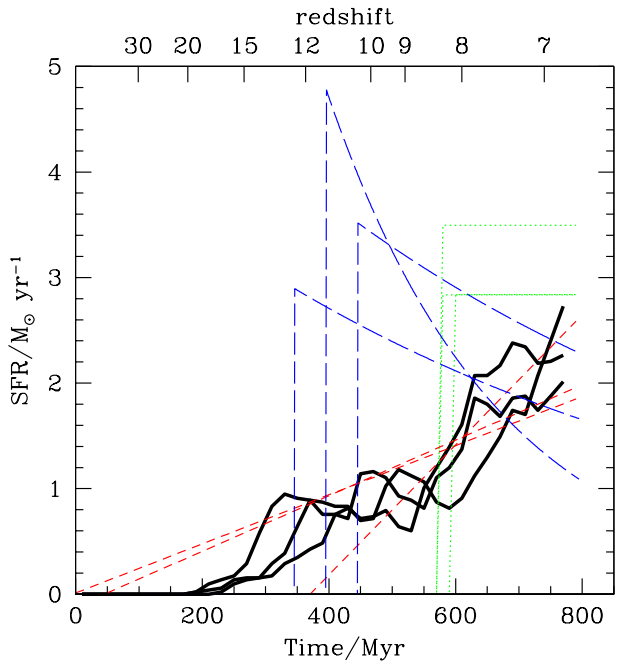}}
\vskip -0.0in
\caption{Star formation histories of the 3 best-fitting vzw galaxies for
Abell 2218 KESR, along with the 3 best-fitting galaxies from each of 
our one-parameter model SFHs.  The SEDs of all models match the data with $\chi^2$
per degree of freedom less than unity.  The vzw SFHs have been sampled
in 20-Myr bins and smoothed with a 100-Myr tophat for readability.  The 
areas under the curves are slightly different, reflecting uncertainty in 
the total stellar mass.  Note that the best-fitting rising model is very 
similar to the best-fitting simulated model. 
}
\label{fig:sfh}
\end{figure}

To date, efforts to use SED-fitting to infer the physical properties
of high-redshift galaxies have generally employed some combination
of constant, exponentially decaying, and single-burst star formation
histories in order to span the presumed range of possibilities.
In general, it has been found that the stellar mass, SFR, and redshift of
a galaxy can be fairly well-constrained via this technique while the age,
metallicity, and dust extinction cannot.  Much of the gain in precision
that results from using simulated galaxies in SED-fitting results from
the relatively small range of SFHs that actually occur in the simulations.

For example, the solid black curves in Figure~\ref{fig:sfh} show the SFHs
of the 3 galaxies from the vzw simulation that yield the best fits to the
$z\sim6.7$ galaxy Abell 2218 KESR, which we will discuss extensively in
\S\ref{sec:kesr}.  The SFHs have been sampled in 20-Myr bins and
smoothed with a 100-Myr tophat in order to make the plot more readable.
All 3 galaxies begin forming stars at $z>15$ and exhibit a SFR that is
generally rising.  An examination of simulated SFHs at these redshifts
shows that steadily rising SFHs are typical.  For this reason we
consider a constantly-rising model SFH in this work; as we will see,
the constantly-rising model reproduces most closely the constraints
obtained from the simulated galaxies (Figure~\ref{fig:degeneracy}).
This model has to our knowledge not been investigated before.

In order to facilitate comparison with much of the available SED-fitting
work that is available in the literature, we investigate three
one-parameter model SFHs for each galaxy in addition to simulated SFHs,
as described below:

\begin{itemize}

\item \emph{Exponentially Decaying SFR} We generate models with SFR proportional
to $e^{-t/\tau}$.  We use four values of $\tau$ logarithmically spaced between
10 and 795 Myr, roughly the age of the universe for our most distant object.
Each of these SFHs is sampled at 23 ages $t$ evenly spaced between 10 and 1000 
Myr.  

\item \emph{Constant SFR} We generate models that have been forming stars at
a constant rate $\sfr$ for $t$ Myr.  For $t$ we sample 41 ages that lie 
between 10 and 1000 Myr, and for $\sfr$ we sample 45 SFRs that lie between 0.2 
and 30.0 $\smyr$.

\item \emph{Constantly Rising SFR} In the constantly rising SFH, a galaxy's 
SFR is proportional to its age.  While a rising SFH can clearly not be 
maintained for all galaxies until low redshifts, it arises fairly 
generically for high-redshift
galaxies in hydrodynamic simulations~\citep{fin06}.  We generate models in
which each galaxy's SFR has been rising at a constant rate for $t$ Myr,
where for $t$ we have sampled 41 ages that lie between 10 and 1000 Myr.

\end{itemize}

For each star formation history, we have generated models with masses in
the range $ \log(M_*/\msun) \in [7.5, 10.5] $ and metallicities $Z_*/Z_\odot
\in (0.005, 0.07, 1.0, 2.5)$.  These SFHs are then put through the
\spoc\ formalism, in order to determine the probability distribution
of physical properties.  During the fitting, we require that the oldest
star of a given model is not older than the age of the universe at the
model's redshift.

\section{Performance of \spoc} \label{sec:performance}
\subsection{Self-Consistency Test} \label{sec:consistency}

We begin by testing that \spoc\ recovers the (known) properties of
simulated galaxies.  This serves to both test the algorithm and quantify
its intrinsic uncertainties.  To do so, we take the 73 galaxies that
are resolved by our vzw simulation at $z=6.5$, and determine how
accurately we can recover their intrinsic physical properties using
model galaxies from the $z=6$ and $z=7$ outputs as inputs to \spoc.
While the model and sample galaxies are not strictly independent in
this test (all but the least massive galaxies at $z=6.5$ correspond to
at least one ancestor in the $z=7$ output and descendant in the $z=6$
output), the galaxies are evolving rapidly enough that these populations
are effectively independent.  The test-case and model galaxies are
compared in 6 bands from $i$-band to IRAC 4.5$\mu$m (the same ones
applied to Abell 2218 KESR in \S\ref{sec:kesr}), where we assume a 0.15
magnitude uncertainty in each band.  The test-case galaxies are reddened
with a fiducial dust extinction $\Av = 0.6$ via the~\citet{cal00} law.
We apply \spoc~to these test-cases using each of the different extinction
curves mentioned in \S\ref{sec:simulations} in order to investigate the
systematic uncertainties resulting from our ignorance of the appropriate
extinction curve for high-redshift galaxies.  During the fitting, redshift
space is sampled by perturbing each model galaxy over a grid extending
to $\Delta z = 0.5$ so that we sample the range $z \in [5.5,7.5]$; $\Av$ 
is sampled over the range $\Av \in [0,1]$.

\spoc\ constrains six quantities: $M_*$, SFR, $A_V$, $Z_*$, age, and
redshift.  The definitions of $M_*$ and redshift are self-evident.  $A_V$
is defined in terms of the \citet{cal00} reddening presciption.  For the
purposes of this work, metallicity $Z_*$ is defined as the mean mass
fraction of metals in the galaxy's stars; this is useful in determining
what metallicity to choose during population synthesis modeling.
Although metallicity is not the dominant factor in determining a galaxy's
SED, the fact that the vzw model reproduces the mass-metallicity
relation of star-forming galaxies at $z\sim2$~\citep{erb06,dav06b}
as well as for the host galaxy of GRB050904, which is located at
$z=6.295$~\citep{ber06,kaw06}, leads us to believe that this model's
predictions for the metallicities of observed reionization-epoch galaxies
are plausible (Finlator et al.\ 2007, in prep.).  We define a galaxy's
age as the mass-weighted mean age of its star particles; this is more
meaningful than the more commonly-used age of the oldest star, which is
both difficult to constrain observationally and difficult to predict owing
to the stochastic nature of our simulations' star formation prescription.
We define a galaxy's SFR as the average over the last 100 Myr leading
up to the epoch of observation; if none of a galaxy's stellar mass is
older than 100 Myr then the age of the oldest star is used.  This metric
is found to correlate more tightly with rest-frame UV flux than averages
over a shorter time-baseline for the numerically simulated SFHs.

\begin{figure}
\setlength{\epsfxsize}{0.5\textwidth}
\centerline{\epsfbox{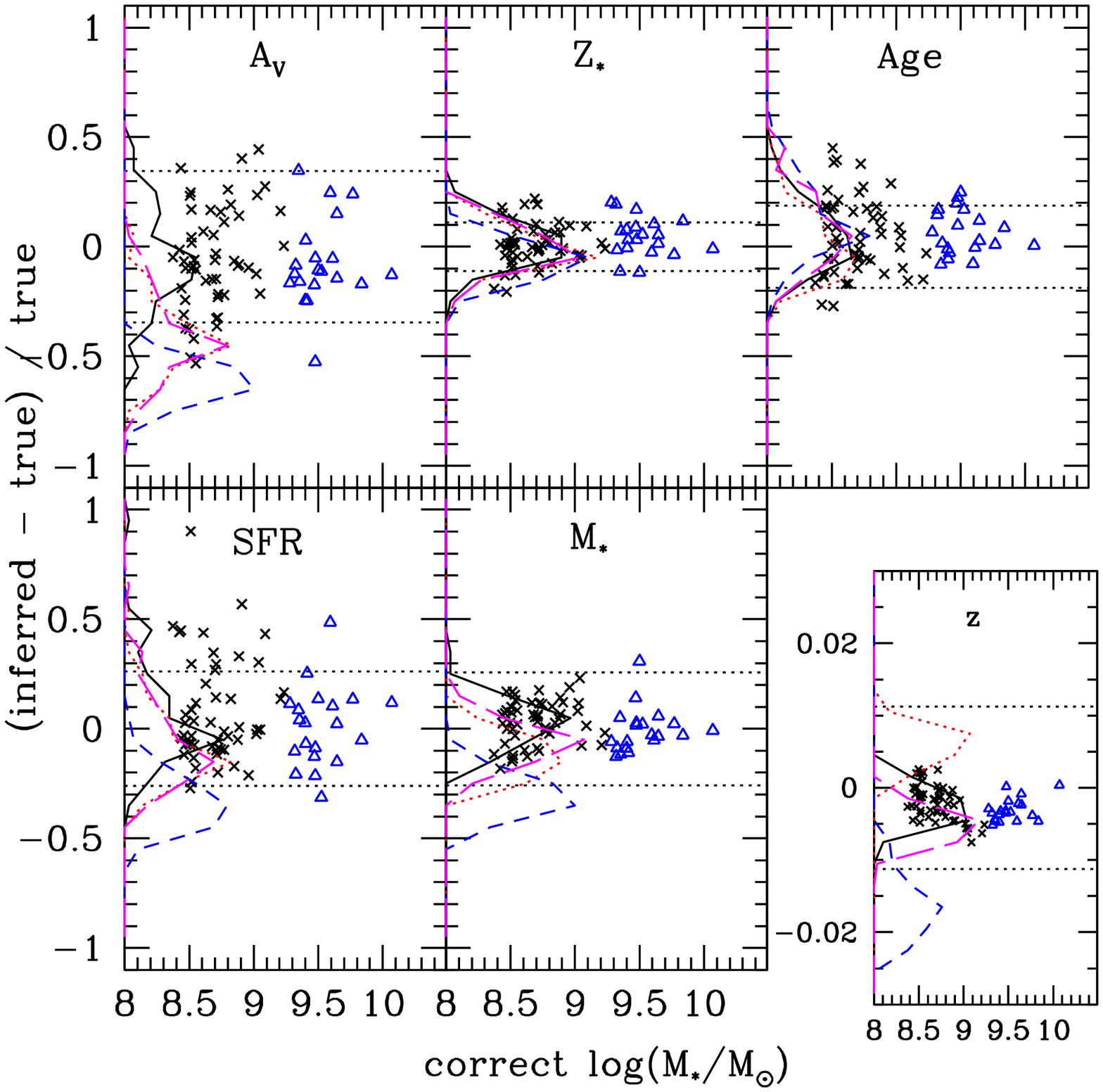}}
\vskip -0.0in
\caption{Fractional error in inferred physical properties of $z=6.5$ vzw
galaxies as determined using the $z=6.0$ and $z=7.0$ galaxies as models
in \spoc.  First we considered the case in which both test-case and model 
galaxies are reddened via the~\citet{cal00} extinction curve.  For this run, 
black crosses and blue triangles denote test galaxies from the $16\hmpc$ and 
$32\hmpc$ volume simulations, respectively; the solid black histogram denotes 
their combined distribution; and dotted lines indicate mean fractional 
$1\sigma$ uncertainties computed by \spoc.  We then compared the same 
test-case galaxies (reddened with the same extinctinction curve) to models 
with the same SFHs but using different dust laws.  From these runs, the dotted 
red, short-dashed blue, and long-dashed magenta histograms correspond to the
cases where the models were reddened with the~\citet{ccm89},~\citet{gor03}, 
and~\citet{cf00} laws.  With few exceptions, the best-fit values are 
within 50\% and $2\sigma$ of the correct values.  Stellar mass and redshift 
are recovered remarkably accurately. 
}
\label{fig:consistency_frac}
\end{figure}
\vskip 0.0in

First we consider the case in which the test-case and model galaxies are 
both reddened via the~\citet{cal00} extinction curve.  For this case, the 
points in Figure~\ref{fig:consistency_frac} show how the fractional error 
in the six inferred properties varies with stellar mass and the solid black
histogram gives their combined distribution.  The dotted lines indicate 
the mean formal $1\sigma$ uncertainties; these are computed directly from 
the probability densities that are returned by \spoc~rather than from 
the scatter in the points.  In general, the recovered physical parameters 
lie within 50\% and $2\sigma$ of the correct values, suggesting that our 
SED-fitting technique is indeed self-consistent.  The fact that the formal 
uncertainties are at least as large as the scatter (and, in some cases, are 
somewhat larger) suggests that the formal uncertainties are sufficiently 
conservative. 

The most accurately (and precisely) recovered parameter is redshift.  The
high accuracy in this case owes to the fact that the $I_{814}$, 
$z_{850}$, and $J_{110}$ fluxes tightly contrain the position of the 
Lyman break, which itself results from the~\citet{mad95} prescription 
for IGM absorption.

Stellar mass is recovered with 20\% accuracy, owing primarily to the
fact that the rest-frame optical flux is generally dominated by numerous
long-lived, low-mass stars whose mass-to-light ratio is relatively
insensitive to age and dust extinction.  Additionally, as we will show
in Figure~\ref{fig:consist_compare_models}, the lack of a significant
systematic offset in the recovered stellar masses owes to the similarity
between the SFHs of the test-case and model galaxies.

Metallicity is also accurately recovered.  This is expected given
that there is a tight mass-metallicity relation in the simulations
(the $1\sigma$ scatter is 15\%) that does not vary strongly with
redshift~\citep{dav06a,dav06b}, and the fact that the test galaxies
and models came from the same simulations.  Without this implicit
prior, metallicity cannot be tightly constrained from broadband
photometry~\citep{pap01,sch05}.

Turning to SFR, we expect a reasonably accurately inferred SFR given 
the tight correlation between SFR and stellar mass that the simulated 
galaxies obey~\citep{fin06,dav06a}; in other words, if the redshift is
known and the stellar mass can be constrained from the rest-frame optical 
flux, then the SFR is already constrained to within a factor of two
regardless of the rest-frame UV flux.  Figure~\ref{fig:consistency_frac}
bears this out.  In detail, SFR is somewhat less accurately recovered than 
stellar mass owing to the degeneracies with age and $\Av$---in fact, a close 
inspection reveals that galaxies with underestimated SFR have overestimated 
$\Av$ and vice-versa.

Age is accurately recovered owing largely to the small range of SFHs that
occur in our simulations.  Just as only a small range of metallicities
remains available once the stellar mass is constrained, a relatively
small range of ages is available once the redshift and stellar mass are
constrained (Figure~\ref{fig:sfh}).

If we relax the assumption that we know the correct form of the dust
extinction curve, we find systematic effects up to the $\approx50\%$ level.  
In Figure~\ref{fig:consist_compare_models}, the dotted red, short-dashed 
blue, and long-dashed magenta histograms correspond to the cases where the 
models were reddened with the~\citet{ccm89}, SMC bar~\citep{gor03}, 
and~\citet{cf00} laws while the test-cases were reddened with 
the~\citet{cal00} law as before.  Metallicity and age are not strongly 
affected because these are tightly constrained by the combination of 
stellar mass and redshift.  In contrast, $\Av$, $\mstar$, and SFR are 
underestimated for the other curves by up to 60\% while the photometric
redshifts are systematically off by up to 2\%, with the the SMC law
yielding the largest underestimates.  These discrepancies owe to the varying 
slopes of the extinction curves: Steeper extinction curves require 
less overall dust (i.e., lower $\Av$) and redder rest-frame UV colors 
(i.e., lower SFR, and thus lower stellar mass in our simulations) in order 
to match a given observed rest-frame UV color.  Similarly, photometric 
redshifts are systematically off because the extra suppression of rest-frame 
UV flux that results from an overly steep extinction curve can be partially
cancelled out by underestimating the galaxy's redshift.

In summary, stellar mass, metallicity, age, and SFR can simultaneously
be recovered by \spoc~when numerically simulated models are used owing
to the existence of implicit priors on these parameters.  Any remaining
discrepancy between the observed and model UV fluxes is minimized by
the choice of $\Av$, which is also relatively accurately recovered.
Thus, our SED-fitting technique is indeed self-consistent.  However,
if the slope of the assumed dust extinction curve is incorrect then the
resulting best estimates of the physical parameters may be off by up to
$\approx 50\%$ while the photometric redshift may be off by up to $2\%$.
These systematic uncertainties are generic to studies of high-redshift
galaxies that employ SED-fitting and are unrelated to uncertainties that
result from our ignorance of the correct form of high-redshift SFHs.
Since it is the latter aspect that we are currently trying to constrain,
we do not further consider SED-fitting errors.

\subsection{Comparison With One-Parameter Models} \label{sec:consist_compare_models}
\begin{figure}
\setlength{\epsfxsize}{0.5\textwidth}
\centerline{\epsfbox{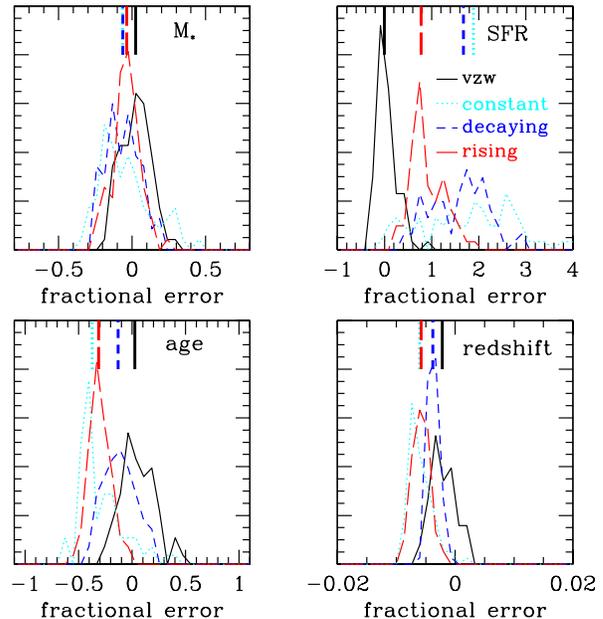}}
\vskip -0.0in
\caption{Comparison of the distribution of fractional
errors in the inferred physical properties of the galaxies from
Figure~\ref{fig:consistency_frac} when assuming different SFHs in \spoc.
Redshift error is plotted as $\delta z /(1+z)$.  All histograms in a
given plot have been normalized to enclose a constant area.  Solid black,
dotted cyan, short-dashed blue, and long-dashed red curves give the
histograms for the simulated, constant, decaying, and rising model sets;
the vertical tickmarks at the top indicate the respective medians.
Redshift is very well-recovered for all models, stellar mass and age
to within 50\%, while SFR can be off by large amounts. 
}
\label{fig:consist_compare_models}
\end{figure}
\vskip 0.0in

It is reassuring but not terribly surprising that \spoc\ can accurately
recover the physical properties of the galaxies that it uses as templates.
A more interesting question is how well \spoc\ can recover galaxy
properties using a different SFH than that of the input galaxy, as
this illustrates the variations in inferred physical parameters among
various assumed SFHs.  To address this, we have fit the test-case
galaxies that were used in \S~\ref{sec:consistency} using model
sets generated from constant, decaying, and rising SFHs as described
in \S~\ref{sec:SFHs}.

Figure~\ref{fig:consist_compare_models} gives the distributions
of fractional errors in the inferred values of stellar mass,
SFR, age, and redshift that result when using the different model
sets.  The vzw case is simply a vertically-binned histogram from
Figure~\ref{fig:consistency_frac}.  Generally, the one-parameter models
yield stellar mass and age results that are within 50\% of the correct
values.  The errors for these quantities are generally distributed with
slightly larger scatter than the errors from the vzw models and show
systematic discrepancies up to the 40\% level.  The SFRs are overestimated
systematically by 50--200\% with significantly more scatter than returned
by the vzw models; this is clearly the quantity that is most dependent on
the assumed SFH.  The vzw models systematically underestimate redshift by
0.2\% while the one-parameter models are low by 0.5\%; the scatters are
comparable for all of the models.  We briefly discuss results specific
to each one-parameter model in turn.

When considering all test-case galaxies together, the constant-SFR
models tend to underestimate the age and stellar mass by 40\% and 10\%
respectively while overestimating the SFR by a median factor of 3,
the largest discrepancy among the SFHs that we consider.  When we
split the sample into ``massive" and ``low-mass" galaxies at $M_*/\msun
= 10^9$, we find that the constant models tend to overestimate the ages
of ``massive" galaxies by $\sim20$\% while underestimating the ages of
low-mass galaxies by $\sim40$\%.  In order to match the rest-frame optical
measurements, the constant-SFR models then overestimate the SFR for the
low-mass and massive galaxies by 100--200\% and 0--100\%, respectively,
with 50\% scatter in each case.  Stellar masses are underestimated by 20\%
for the low-mass galaxies and overestimated by roughly the same amount
for massive galaxies.  This illustrates that uncertainties in parameter
recovery are not only dependent on the assumed SFH, but also on the mass.

The decaying models tend to reproduce the stellar mass and age with
systematic errors of roughly 10\% and scatter comparable to the scatter
from the vzw models.  These successes are somewhat surprising because
the simulated SFHs look nothing like the decaying case.  Conversely,
the SFRs are higher by a median factor of 2.8, only slightly better
than the constant model.  The fact that the SFR could be dramatically
overestimated while the age, stellar mass, and dust reddening (not shown)
are recovered accurately probably owes to our use of 100-Myr average SFRs.
When considering all of the stellar mass that has formed in the last 100
Myr, a larger fraction of the O-stars will have evolved off of the main
sequence for decaying or constant SFHs than for rising models that show
the same 100 Myr average SFR, leading to the result that models with
differing SFRs nonetheless produce similar UV luminosities.  The broad
success of the decaying model despite the input SFHs looking nothing
like the decaying case shows that SED fitting can yield interpretations
consistent with monolithic collapse models even though the true SFH may
be quite different.

The rising models tend to underestimate the age by 10--40\% and
overestimate the SFR by 40\%, though with significant scatter.  Both of
these offsets are compensated by overestimated $\Av$ in such a way
that the stellar masses are recovered quite accurately, with $<5$\%
systematic offset and scatter comparable to what is achieved via the
numerically-simulated models.  Overall, this model probably recovers
the true parameters most faithfully among the one-parameter models,
though it is not a dramatic improvement over the others.

In summary, we have shown that \spoc\ can self-consistently recover
the physical properties of the model galaxies that we use in fitting
observed high-redshift galaxies.  Further, simple one-parameter models
are able to recover stellar mass and age to within 50\% accuracy and SFR
to within a factor of three, although there are systematic offsets at
a comparable level.  All models yield photometric redshifts with better
than 1\% accuracy although none of them outperform the numerical models.

\subsection{Numerical Resolution} \label{sec:resolution}

\begin{figure}
\setlength{\epsfxsize}{0.5\textwidth}
\centerline{\epsfbox{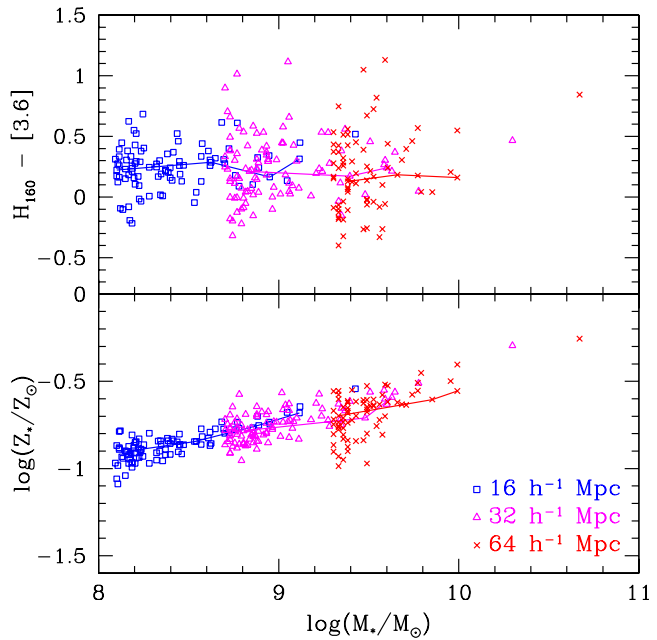}}
\vskip -0.0in
\caption{Numerical resolution convergence test I.  (Top) Rest-frame UV-optical
color versus stellar mass at $z=6$.  The blue squares, magenta triangles, 
and red crosses correspond to the resolved galaxies from the 16, 32, and 64 
$\hmpc$ simulations, respectively; running medians are also given.  (Bottom) 
Stellar metallicity versus stellar mass for the same galaxies.  In both cases,
the median trend and the scatter do not vary with scale, indicating that the
SFHs of our simulated galaxies do not suffer from numerical resolution issues.
}
\label{fig:catalog_resolution}
\end{figure}

\begin{figure}
\setlength{\epsfxsize}{0.5\textwidth}
\centerline{\epsfbox{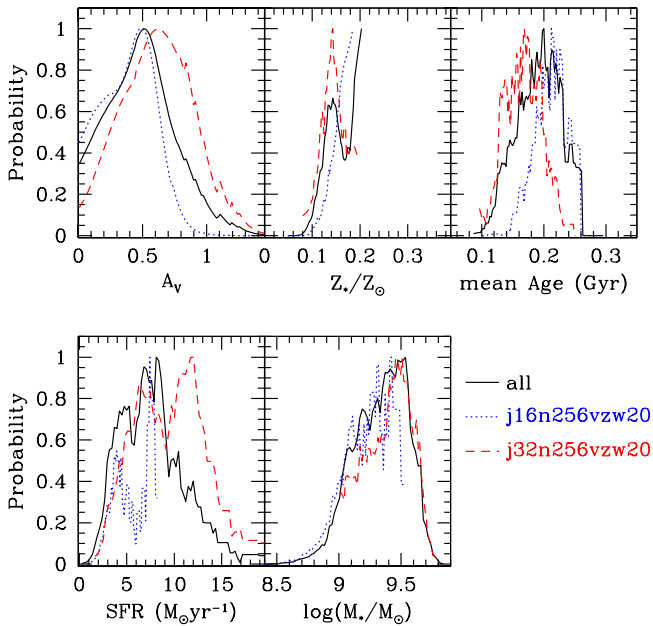}}
\vskip -0.0in
\caption{Numerical resolution convergence test II.  The dotted blue and 
dashed red curves give the probability densities for
the physical properties of A370 HCM6A as derived from model galaxies from
a $16\hmpc$ and a $32\hmpc$ simulation volume, respectively, while the solid
black curve results from combining the models.  All of the probability
densities show agreement at the $1\sigma$ level and only the age curves
show clear evidence of resolution effects (see text).  This figure 
indicates that our inferred physical properties are not significantly 
hampered by numerical resolution issues.
}
\label{fig:spoc_resolution}
\end{figure}

The \spoc\ library of simulated galaxies that we employ is accumulated
from simulations at different volumes and resolutions (e.g. in the jvzw
case, we use the $16, 32, 64\hmpc$ runs).  In \citet{dav06a}, we showed
that, down to the adopted stellar mass resolution limit, the physical
properties of galaxies are similar at overlapping mass scales between
the various simulations.  To reiterate this point in a way that is more
relevant to the current work, Figure~\ref{fig:catalog_resolution} shows
how the rest-frame UV-optical color and mean stellar metallicity vary with
stellar mass at $z=6$ for resolved ($\geq 64$ star particles) galaxies
from our 16, 32, and 64 $\hmpc$ simulation volumes.  In both cases,
the trend and the scatter are consistent between the different volumes,
giving us further confidence that our simulated SFHs are in fact resolved.

A related way to look for resolution issues is to ask whether \spoc\
will yield similar answers when it is applied to same-mass galaxies
at different resolutions.  We now demonstrate that, indeed, \spoc\
does recover similar parameters for galaxies at different resolutions,
and so combining different resolution simulations into a larger set is
justified.  Of course, one does not need to do so in order to use \spoc,
it is merely a convenient avenue to increase the dynamic range spanned
by our model galaxies.

For convenience, we study A370 HCM6a as it can readily be fit by our 
models.  We apply \spoc\ using three sets of models derived from the jvzw
simulations: once using models from the $16 \hmpc$ volume (``j16"),
once using models from the $32 \hmpc$ volume (``j32"), and once using
both sets.  A lack of numerical convergence would result in systematic
offsets between the probability densities from the first two fits,
while the combined result shows how the models from the two volumes
combine to yield our full probability density.

Figure~\ref{fig:spoc_resolution} shows the derived probability density
functions for the various physical parameters that we consider.
The ranges agree well, and the best estimates from the j16 and j32
volumes (defined as the means of the probability density functions)
are consistent at the $1\sigma$ level.  In detail, the j16 models return
fits with somewhat lower stellar mass, SFR, and metallicity than the j32
models while the j32 models yield ages that are younger by about 40 Myr.
The small offsets in mass, SFR, and metallicity do not indicate resolution
problems as they are expected even in the absence of any convergence
issues.  Briefly, the j16 volume contributes lower mass models owing
to the slope of the mass function (at the massive end) and our 64 star
particle mass resolution cut (at the low-mass end).  On the other hand,
the age offset results from a well-known numerical resolution limitation
whereby galaxies in lower-resolution simulations take longer to condense
beyond a given critical density in order to begin forming stars, yielding
younger ages at a given stellar mass and redshift.  Fortunately, the
offset is comparable to the intrinsic uncertainty on this parameter.
We have repeated this test using object SBM03\#1 with the 32 and $64
\hmpc$ volumes and found similar results.  Hence we do not believe that
our results using combined simulation samples are significantly hampered
by numerical resolution effects.

\section{Test Case: Abell 2218 KESR} \label{sec:kesr}

\begin{figure}
\setlength{\epsfxsize}{0.5\textwidth}
\centerline{\epsfbox{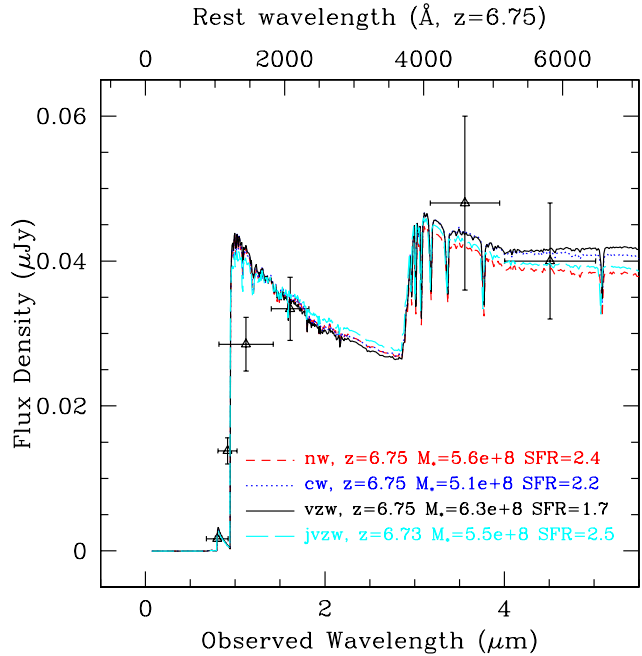}}
\vskip -0.1in
\caption{Best-fit spectra from the simulations to the data for Abell
2218 KESR, which we have demagnified by $25\times$~\citep{kne04}.
Spectra from the nw, cw, and vzw, and jvzw models are denoted by red 
dashed, blue dotted, black solid, and long-dashed cyan curves, 
respectively.  All four models produce galaxies whose spectra match 
Abell 2218 KESR within the errors ($\chi^2$ per degree of freedom $<1$). 
}
\label{fig:spectra}
\end{figure}
\vskip 0.3in

The triple arc in Abell 2218, dubbed Abell 2218 KESR by~\citet{sch05}
after its discoverers~\citep{kne04}, is probably the best-studied $z>6$
object at present, and its physical parameters have been constrained
through SED fitting by various authors.  Hence it provides a good test
case for exploring the systematics that result from using numerically
simulated model galaxies, and comparing to results employing more
traditional simple SFHs.

The flux from Abell 2218 KESR can be measured from two lensed images in the
\emph{Hubble} Advanced Camera for Surveys (ACS) $z_{850}$, Wide-Field 
Planetary Camera 2 (WFPC2) $I_{814}$, and Near-Infrared Camera and 
Multi-Object Spectrograph (NICMOS) $J_{110}$ and $H_{160}$ bands, and 
only one image in the \emph{Spitzer}/IRAC 3.6 and 4.5~$\mu$m bands 
(the other image is blended with a nearby submillimeter source 
at IRAC's spatial  resolution).  \citet{sch05} note that the fluxes measured 
by different authors in the optical/near-infrared bands disagree due to the
inherent difficulties of measuring photometry from extended arcs, and that
the different images of the galaxy do not agree in the \emph{Hubble}/ACS
$z_{850}$ band.  Following their suggestion, we use the weighted mean
of the two images in the optical/near-infrared bands (their SED1) and
impose a minimum 0.15~mag uncertainty in all bands in order to account
for differential lensing across the images.  \footnote{\citet{sch05} have
noted that the published upper limits from LRIS and in the \emph{Hubble}/ACS
$V_{606}$ band do not significantly affect the derived parameters.
In the case of the $V_{606}$ limit, we have verified this.}

\begin{figure}
\setlength{\epsfxsize}{0.5\textwidth}
\centerline{\epsfbox{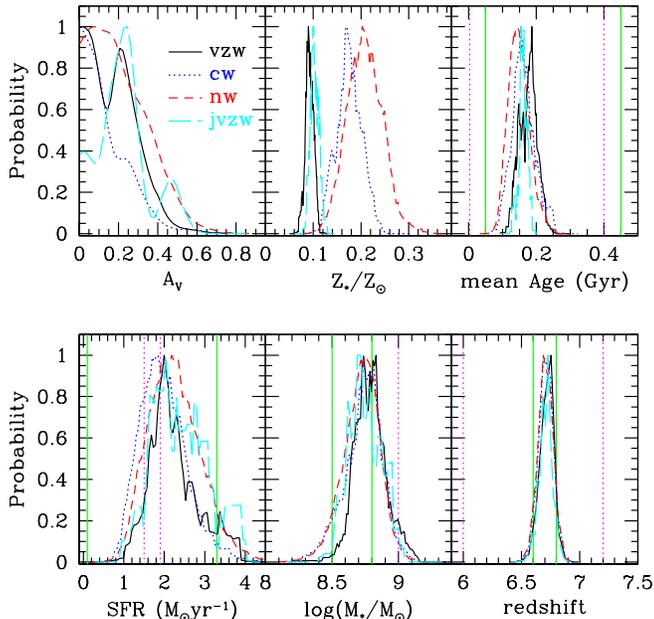}}
\vskip -0.2in
\caption{Probability densities for the physical parameters of Abell 2218
KESR based on four numerical simulations assuming different treatments
for superwind feedback.  Results from the nw, cw, vzw, and jvzw models are
given with dashed red, dotted blue, solid black, and long-dashed cyan 
lines, respectively.  Solid green and dotted magenta boundaries denote
ranges inferred by~\citet{ega05} and~\citet{sch05}, respectively (the
lower limit to stellar mass obtained by~\citet{sch05} is 7.7).
Each curve has been normalized to unit area and then scaled so that
the three models fit on the same plot.  All of the best-fit physical
parameters except metallicity (see text) are robust to changes in the
superwind feedback treatment. 
}
\label{fig:compareWind}
\end{figure}

\subsection{Modeling Uncertainties: Outflows and Dust}

\label{sec:windDust} Figure~\ref{fig:spectra} shows that the SEDs of the
best-fitting model galaxies from our three galactic outflow recipes and
two cosmologies all reproduce the observations with reduced $\chi ^2<1$.
Moreover, they are remarkably similar.  All four models possess a very
blue rest-frame UV continuum owing to young age and low metallicity as
well as a pronounced Balmer break owing to the presence of older stars.
The best-fit parameters are similar, indicating that the simulation's
ability to reproduce the observations is robust to the choice of
superwind feedback prescription and detailed cosmological parameters
(to the extent of the variations considered).  This result is akin to
the findings from studies employing one-parameter model SFHs that good
fits can be obtained via a variety of assumed SFHs and metallicities.

To quantify this point, Figure~\ref{fig:compareWind} and 
Table~\ref{table:compareWind} show how the derived probability
densities for the physical parameters of Abell 2218 KESR depend on the
choice of wind model.  The entries in Table~\ref{table:compareWind} list the
mean and variance of the histograms shown in Figure~\ref{fig:compareWind}.
Each curve in Figure~\ref{fig:compareWind} has been normalized to unit
area and then scaled so that all four curves fit on the same plot.

All of the derived physical parameters except $\zstar$ are remarkably
robust to our choice of superwind feedback prescription.  Generally,
the data seem consistent with negligible dust reddening, a mean
stellar age of 100--200 Myr, SFRs of 1--3 $\smyr$, a stellar mass of
3--8$\times 10^8 \msun$, and a redshift $z\sim6.7$, in good agreement
with other determinations~\citep{ega05, sch05}\footnote{Note that
because our simulations' star formation treatment assumes instantaneous
recyling, we give here the total mass of stars that have formed.
Using the~\citet{bru03} tables, we find that $\approx70$\% of the stellar
mass of stars formed in galaxies that are more massive than $10^8 \msun$
at $z=6.75$ remains in stellar form at $z=6.75$.}. The tightness of the
constraints on the various intrinsic physical parameters results from
the relatively narrow range of SFHs experienced by the galaxies in the
simulations.  By contrast, the uncertainty in the inferred redshift is
determined by our treatment of IGM extinction since the inferred redshift
is dominated by the position of the Lyman-$\alpha$ break.

Comparing the results from the models in detail, both the cw and nw models
are less efficient at suppressing star formation via outflows.
For this reason, at fixed number density (or equivalently, dark matter
halo mass) the cw and nw model galaxies have formed more stars, retained
more of their metals, have a higher SFR, and are older~\citep{dav06a}.
Similarly, at a given stellar mass, the vzw galaxies are younger,
have expelled a larger fraction of their heavy elements, and for these
reasons require more dust reddening in order to match a given observed
colour owing to the age--metallicity--dust degeneracy.  The jvzw models
give results that are the similar to the vzw results owing to the 
similar feedback treatment.  However, the lower values used in the jvzw
simulation for the cosmological parameters $\Omega_m$ and $\sigma_8$ 
delay the growth of structure at early times, yielding fewer model 
galaxies for us to match against observations for a given simulated 
volume.  As a result, parameter space is less well-sampled and the 
probability density curves (notably $\Av$) exhibit more stochasticity.

Next to stellar mass and redshift, the mean stellar metallicity is the
most tightly constrained parameter, followed by the mean age and SFR.
Metallicity is the only parameter for which the different wind models
disagree at the $>1\sigma$ level.  That the models could agree on all
of the derived parameters except for the metallicity reflects the fact
that the effect of metallicity on the photometry is small compared to the
effect of stellar age~\citep{sch05}.  On the other hand, the tightness
of the metallicity constraint follows from the tight mass-metallicity
relation that the simulation predicts (\S~\ref{sec:performance}).  Thus,
the apparent disagreement between the metallicity constraints is simply
a reflection of the tight priors imposed by the different simulations
combined with the robust constraints on stellar mass.  Note that
because the vzw model agrees the best with the distribution of metals
in the IGM~(OD06) and the mass-metallicity relationship of star-forming
galaxies~\citep{dav06b}, it probably makes the most believable prediction
of Abell 2218 KESR's mean stellar metallicity.

We investigated the effect that varying the dust prescription has on the
derived physical parameters and found, in agreement with~\citet{sch05},
that this has no significant effect on the derived physical parameters
other than $\Av$ when the simulated models were used.  The derived $\Av$
were roughly 0.1 mag lower for the~\citet{ccm89} and~\citet{cf00} laws;
these differences are expected given the extra extinction imposed at
rest-frame UV wavelengths in the former case and (similarly) for younger
stars in the latter case.  The total amount of light removed by dust
extinction, which can be regarded as a prediction of the total infrared
luminosity, is roughly $10^{10} L_{\odot}$, independent of the assumed
dust prescription.

Two groups have previously published constraints on this object's
properties.  \citet{ega05} employed a uniform sampling of single-parameter
SFHs; their results are given in Table~\ref{table:compareWind} and by
the solid green vertical lines in Figure~\ref{fig:compareWind}, where
we have converted their derived SFR and $\mstar$ to values appropriate
for a Chabrier IMF.  It is clear that their constraints are entirely
consistent with our own, although the tight intrinsic correlations
between physical parameters in the simulated models allow us to impose
tighter constraints on all of the derived parameters.  In the second
work,~\citet{sch05} exhaustively examined the systematics of assumptions
regarding SFH and template spectra with the goal of bracketing the most
likely parameter space.  Their constraints, given by the dotted magenta
vertical lines and listed in~\ref{table:compareWind}, are also consistent
with ours.  Hence our simulations, despite a different generic form for
their galaxies' SFHs than has been assumed in previous investigations,
can reproduce the properties of Abell 2218 KESR equivalently well.

\subsection{Comparison to One-parameter SFHs} \label{sec:sim_vs_1param}

\begin{figure}
\setlength{\epsfxsize}{0.5\textwidth}
\centerline{\epsfbox{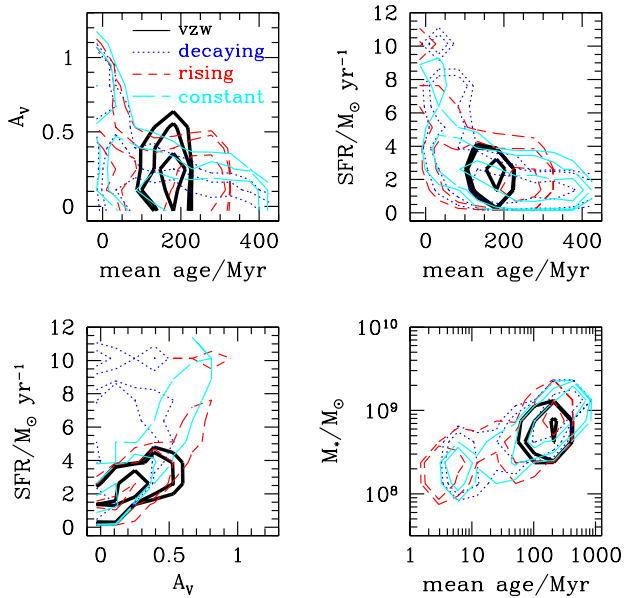}}
\vskip -0.0in
\caption{Best-fit combinations of age, extinction, and and SFR for
1000 Monte Carlo re-observations of Abell 2218 KESR.  The solid black,
dotted blue, short-dashed red, and long-dashed cyan contours enclose
68\%, 95\%, and 99\% of the best-fit solutions for the vzw, decaying,
rising, and constant model sets, respectively.  The vzw models allow
for the tightest constraints to be placed.  Interestingly, the $1\sigma$
best-fit intervals from the constantly-rising SFH are very close to the
simulation result. 
}
\label{fig:degeneracy}
\end{figure}

Efforts to constrain the physical properties of high-redshift galaxies
via SED-fitting invariably encounter a host of degeneracies between
the best-fit parameters, the most difficult of which is certainly the
age-extinction-SFR degeneracy~\citep{sha01,pap01,sha05}.  Generally, young
stellar population age, high SFR, and low dust extinction all yield bluer
photometric colours while old age, low SFR, and high dust extinction all
yield redder photometric colours.  These degeneracies also contribute to
the uncertainty in the inferred stellar mass via the stellar age-stellar
mass degeneracy, whereby older populations have a higher mass-to-light
ratio, yielding a higher stellar mass at a given measured flux density.
Compounding the problem, there are a number of ways in which the
best-fit physical properties depend on any assumptions that are made
regarding the shape of a galaxy's star formation history~\citep{sha05}.
Numerical simulations, in contrast, provide tight relationships between
galaxy star formation rate, stellar mass, and metallicity (from which
extinction may be inferred).  These predicted relationships translate
into tighter constraints on physical parameters that depend strongly on
the shapes of the trial SFHs.

In order to quantify how effectively the simulations reduce well-known
degeneracies in best-fit parameters, we generated 1000 Monte Carlo
re-observations of Abell 2218 KESR by adding scatter to the photometric 
measurements in a way that was consistent with the reported observational 
uncertainties.  For each data set, we then determined the best-fit parameters 
from the vzw models as well as from the decaying, rising, and constant SFH 
models.  Figure~\ref{fig:degeneracy} shows the locus of best-fit parameters 
in several famously degenerate projections of parameter space.  The solid
black, dotted blue, short-dashed red, and long-dashed cyan contours
enclose 68\%, 95\%, and 99\% of the best-fit solutions for the vzw,
decaying, rising, and constant model sets, respectively.

The expected degeneracies that result from one-parameter SFHs are easy to
see in each panel of Figure~\ref{fig:degeneracy}.  At the young end,
each of the one-parameter models includes a parameter space corresponding to a
galaxy that has formed all of its stars in a rapid burst lasting less
than 100 Myr.  Our use of 100-Myr average SFRs guarantees these models
a high SFR ($\geq 4 \smyr$).  Because these models are very young, they
are intrinsically blue and can require high dust reddening ($\Av \leq 1$
mag) to match the observations.  Additionally, their young ages guarantee
that their observed optical flux is dominated by short-lived O and B stars
with low mass-to-light ratios, leading to low inferred stellar masses.
These dramatic burst-dominated models have no analogue in the simulations.

The old end is dominated by the constant-SFR models, equivalent to
the $\tau\rightarrow\infty$ limit of the decaying models.  These models are
characterized by intrinsically red colours and high mass-to-light ratios,
leading to low dust extinctions and high stellar masses.  The oldest of
these fits requires the galaxy to have been forming stars at $\approx
1 \smyr$ when the universe was less than 10 Myr old; our simulations
cannot produce this because gas densities have not grown high enough to
support such SFRs at such early times.

At the $1\sigma$ confidence level, the one-parameter model that most closely
approximates the simulated galaxies is the rising SFH model, albeit with
a preference for somewhat older ages.  Returning to Figure~\ref{fig:sfh},
we see that this is expected because the simulated galaxies' SFHs are
generically characterized by a slowly rising SFR at these redshifts.
Conversely, at the $\ga 2\sigma$ level, the constantly rising SFH
model allows for a wider range of models that have no analogues in
the simulations.  The relatively small range of simulated galaxy SFHs
typified by the examples in Figure~\ref{fig:sfh} leads directly to
the relatively tight range of inferred ages for the solid contours in
Figure~\ref{fig:degeneracy}.  Assuming that Abell 2218 KESR is located at
$z=6.75$, the SFHs in Figure~\ref{fig:sfh} suggest that it may have formed
its first stars before $z=15$, with (10\%, 50\%) of its stars in place
by $z=$(11,8); in other words, roughly half of the best-fitting models'
stars at the epoch of observation are over 150 Myr old.  This relatively
old population readily accounts for the pronounced Balmer break that is
visible in Figure~\ref{fig:spectra}, and is typical in our simulations.

In summary, a key point of this paper is demonstrated by the fact that,
in each panel of Figure~\ref{fig:degeneracy}, the confidence intervals
obtained from simulated galaxies fall well within the range described by
the complete set of one-parameter model SFHs, while yielding the tightest
constraints.  The tighter constraints owe to the relatively small range
of SFHs and the tight correlations between the various parameters that
generically occur in hierarchical simulations of galaxy formation.
This illustrates that, if the simulations are broadly correct,
the physical properties of high-$z$ galaxies can be more precisely
constrained using \spoc.  However, since at present the simulations
are largely untested at these epochs~\citep[modulo the broad successes
in][]{dav06a}, it may be more appropriate to regard the tight simulation
constraints as predictions to be tested against future observations.

\subsection{The Importance of Rest-Frame Optical Data} \label{sec:opticalData}
\begin{figure}
\setlength{\epsfxsize}{0.5\textwidth}
\centerline{\epsfbox{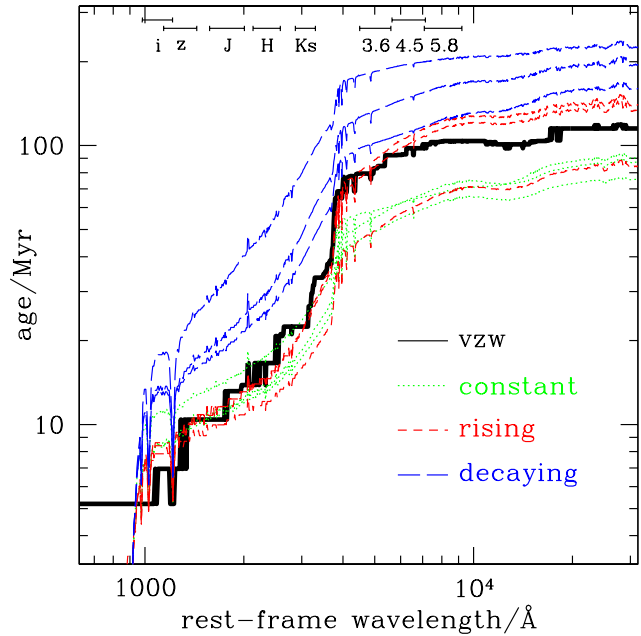}}
\vskip -0.0in
\caption{Light-weighted median age of A2218 KESR versus rest-frame 
wavelength.  The single solid curve corresponds to the best-fitting 
vzw model while the dotted green, short-dashed red, and long-dashed 
blue curves correspond to the three best-fitting constant, rising, 
and decaying models, respectively.  The ranges at the top of the figure
indicate the full width at 20\% of maximum response for an object located 
at $z=6$.  The specific one-parameter models whose 
spectra are considered in Figure~\ref{fig:ffba} are the same as the
models in Figure~\ref{fig:sfh} except that we include only the best-fitting
vzw model for clarity. Generally, measurements shortward of the Balmer 
break sample light from stars that are 50-100 Myr old whereas measurements 
in rest-frame optical wavelengths constrain older populations.
}
\label{fig:ffba} 
\end{figure}
In this work we focus on reionization-epoch objects for which
rest-frame optical measurements are available because the rest-frame optical
flux is dominated by relatively low-mass stars whose mass-to-light ratio
evolves slowly relative to the stars that dominate the rest-frame UV.  
For reasonable SFHs, this allows tighter constraints to be placed not only 
on the galaxy's stellar mass, but also on its SFH.  To quantify this point,
Figure~\ref{fig:ffba} plots the light-weighted median age of A2218 
KESR versus rest-frame wavelength for the SFHs in Figure~\ref{fig:sfh}.  
For example, a point at 2000 \AA~and 15 Myr indicates that, for that model, 
50\% of the photons with rest-frame wavelength of 2000 \AA~are emitted by 
stars that are 15 Myr old or younger.  

For all of the models that we consider, photons from bluewards of
the Balmer break are generated by stars that are under 100 Myr old.
This is expected since B stars live roughly 100 Myr~\citep{ibe67}
and justifies the use of rest-frame UV flux as a constraint on the
current SFR.  Conversely, it explains why rest-frame UV data do not
constrain a galaxy's SFH prior to $\approx 100$ Myr before the epoch
of observation.  By contrast, data from longer wavelengths sample the
SFH at earlier epochs because the stars that dominate these wavelengths
live much longer.  In the case of A2218 KESR, the constant and decaying
models in Figure~\ref{fig:ffba} suggest that rest-frame optical data
constrain the galaxy's SFH roughly 80 and 200 Myr before the epoch
of observation, while the rising and vzw models fall in between these
limits; the differences probably owe to the detailed interplay between
the slope of the SFH and the rate at which low-mass stars fade.
It is possible that the rough agreement in the optical portion of
Figure~\ref{fig:ffba} explains the tendency of all of the best-fit
models in Figure~\ref{fig:sfh} to be in rough agreement with each
other during the interval $z\approx8\rightarrow6.7$ even though they
diverge prior to $z=8$.  Note that this agreement is nontrivial given
that all of the one-parameter models can in principle match the observations with
burst-like solutions (Figure~\ref{fig:degeneracy}).  Most importantly,
it highlights the potential of rest-frame optical measurements redwards
of the Balmer break to constrain the SFH for reionization-epoch galaxies.

\section{A Sample of Reionization-Epoch Galaxies} \label{sec:otherGals}
\begin{figure}
\setlength{\epsfxsize}{0.5\textwidth}
\centerline{\epsfbox{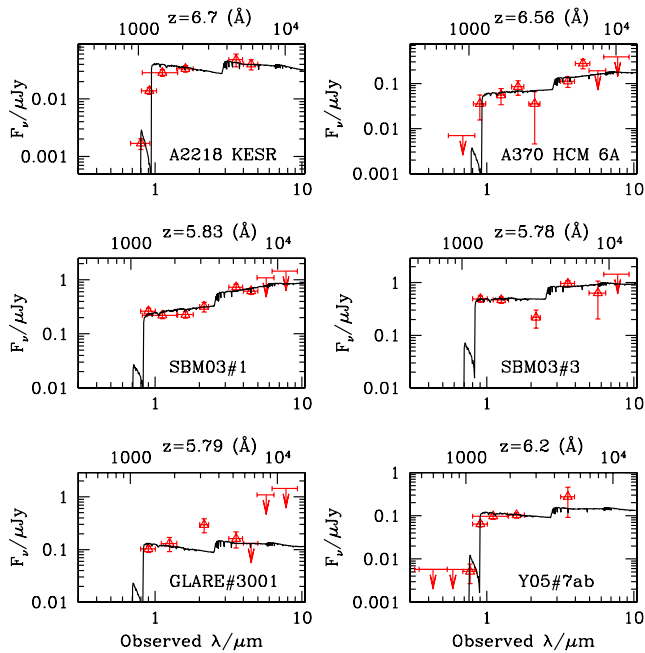}}
\vskip -0.0in
\caption{Spectra of the best-fitting model galaxies overplotted on
the measured photometry.  In cases where spectroscopic redshifts are
available this has been forced; in other cases (Abell 2218 KESR and
Y05\#7ab) we used the photometric redshifts.  Horizontal error bars
indicate the full width of each filter at 20\% of maximum response.
The bottom and top horizontal axes give the observed and rest-frame
wavelengths.  Using the numerically-simulated SFHs with the~\citet{bru03} 
models, we obtain satisfactory fits in all cases and excellent fits 
in 5 cases.
}
\label{fig:many_objs}
\end{figure}
\begin{figure}
\setlength{\epsfxsize}{0.5\textwidth}
\centerline{\epsfbox{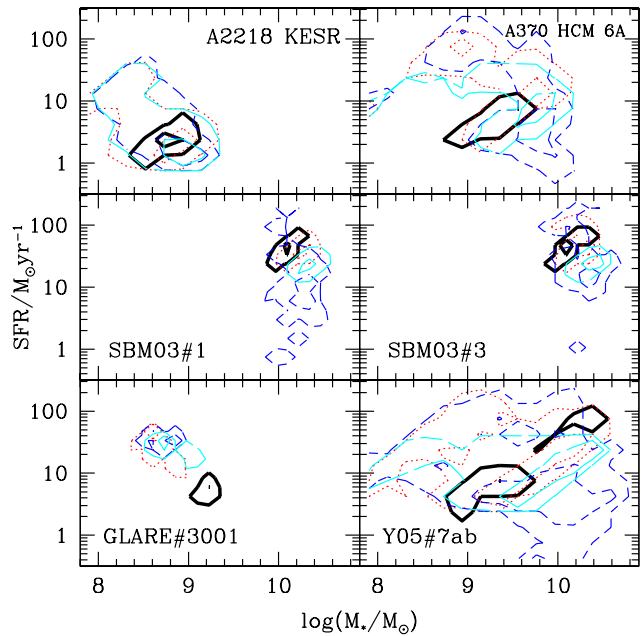}}
\vskip -0.0in
\caption{Contours enclosing 68 and 99\% of the solutions found for the
various galaxies given the various SFHs that we inspected.  Thick solid black,
dotted red, solid cyan, and dashed blue curves correspond to the jvzw,
rising, constant, and decaying SFH models, respectively.  In all cases but
GLARE\#3001, the jvzw models yield results that are generally consistent with
the one-parameter model SFHs while allowing for equally-tight or tighter uncertainties.
}
\label{fig:degAll}
\end{figure}

While the example of Abell 2218 KESR is illustrative, the ultimate
purpose of \spoc\ is to constrain physical parameters for a large
sample of galaxies, in order to characterize the observed galaxy
population and to constrain the underlying galaxy formation model.
To illustrate the sort of insights gained using \spoc, we apply it
to a sample of observed $z>5.5$ objects that have published broadband
photometry in the rest-frame UV and optical bands.  For each object, we
used \spoc~to determine its best-fit physical parameters with the jvzw,
constant, rising, and exponentially-decaying models.  For objects whose
redshift has been measured spectroscopically, we run \spoc~twice: once
with the redshift constrained to the measured value and once with the
redshift left as a free parameter.  The first run allows more accurate
derivations of the physical parameters, while the second enables us 
to study the accuracy of \spoc's photometric redshifts.

Table~\ref{table:many_objs} compares the means and 95\% confidence intervals 
for the various physical parameters of each galaxy that result from the 
various models, while Figure~\ref{fig:degAll} shows the 68 and 99\% 
contours in the $\mstar$-SFR space for the same models.  While fitting with
one-parameter models, we confirmed that the probability density varies only weakly
with $\zstar$~\citep{pap01}; for this reason we have omitted $\zstar$
from Table~\ref{table:many_objs}.  Our simulations, combined with \spoc,
suggest that all of the objects have $0.1 \la \zstar/\zsun \la 0.3$
\citep{dav06a}.

Figure~\ref{fig:many_objs} compares the synthetic spectra of the best-fit
jvzw models with the published rest-frame UV through optical data for our
sample.  Table~\ref{table:many_objs} shows that we obtain very good fits
($\chi_\nu^2 \la 2$) with the simulated library in all cases except for
the unusual object GLARE\#3001.  Note that these spectra and $\chi_\nu^2$
values are purely representative; in practice we obtain physical
constraints through the Bayesian analysis described in \S\ref{sec:spoc}
rather than by simply examining the best-fitting model.  It is encouraging
that the fits obtained from our simulated galaxy library are generally
equally as compelling, in terms of goodness-of-fit measure $\chi_\nu^2$,
as the fits that result from the one-parameter SFH model libraries.

Taken together, these findings strongly support the idea that most
observed reionization-epoch galaxies possess theoretical analogues within
our simulations.  However, some of these objects show minor discrepancies
that may be hinting at some failing in the models.  Given the currently
large observational uncertainties it is difficult to place robust
constraints on the models, but it is nevertheless worthwhile to discuss
each of these systems in detail in order to illustrate how \spoc\ can
be used to both reveal physical characteristics as well as test the
underlying model.

\parname{A370 HCM 6A} We have taken the photometric data and errors in the
observed R,Z,J,H,K', and \emph{Spitzer}/IRAC bands from~\citet[][hereafter
CSE05]{cha05}, and we have adopted their lensing magnification factor
of 4.5.  The measured $4.5\mu$m flux is anomalously large;~CSE05 interpret
the excess over the inferred stellar continuum as H$\alpha$ emission line
flux and derive an SFR of $140\pm90 \smyr$.  In our simulations galaxies
at $z=6.5$ that possess this object's 3.6$\mu$m flux are forming stars at
2--12$\smyr$.  Since such systems are not expected to produce H$\alpha$
equivalent widths that would contribute significantly to the broad-band
flux, we adopt the uncorrected 4.5$\mu$m measurement.  We verified that,
if we use the H$\alpha$-corrected 4.5$\mu$m flux obtained by~CSE05, the
inferred SFR drops from $6.1\pm2.3$ to $4.9\pm1.9\smyr$; this value is
in conflict with the assumed line strength as expected.  The $8.0\mu$m
limits quoted in the text of~CSE05 are different from the limit plotted
in their Figure 2; we adopt the less restrictive limit from the Figure.

When we do not enforce the spectroscopic redshift, the photometric
redshift $5.9\pm0.4$ is dominated by the $R-Z'$ limit and is roughly
$1.5\sigma$ below the spectroscopic redshift; the formal redshift
uncertainty is dominated by the size of the $R-Z'$ baseline.  Although
the one-parameter models yield more accurate photometric redshifts in this case,
the difference is comparable to the formal uncertainty and we regard
it more as a coincidence than as a clue to the SFH of this object.
A more accurate photometric redshift would require us to include the
unpublished measurement in $I$ and an accurate $Z'$ transmission curve;
we have used the SDSS $z'$ profile because a transmission curve for the
$Z'$ measurement used by~\citet{hu02} is not available.

The fits that we obtain are very good, although we confirm that
the 4.5$\mu$m measurement is difficult to interpret as purely
stellar continuum emission.  In fact, even without the anomalous
$4.5\mu$m flux the UV continuum is difficult to understand owing
to the dip at $K'$~\citep{sch05}, and the measured \lya flux is
difficult to reconcile with the observed \lya luminosity function at
$z\sim6.5$~\citep{mal04,kas06}.  Thus, until more precise measurements
of this object are available, any inferences regarding its physical
properties should probably be regarded as suggestive.

With these caveats in mind, we find that using the simulated models
yields inferred stellar mass, SFR, and $\Av$ that are fully consistent
with previous derivations as well as with the one-parameter model SFHs.  The range
of stellar masses allowed by the jvzw models nicely brackets the range
between the emission line-dominated and stellar continuum-dominated
solutions suggested by~CSE05, and the range of SFRs is consistent with
the derivation of~\citet{hu02} based on the \lya line and well below
the inferred H$\alpha$-based SFR of~CSE05.  Turning to the one-parameter models,
the widest uncertainties for all of the parameters are provided by the
decaying model, with the results from the constant and rising models
spanning a somewhat smaller subset of this space.  The results from the
rising model are generally the closest match to the jvzw models although
they permit a somewhat larger range of SFR; these are burst-like models
that have no analogue in our simulations.

\parname{SBM03\#1, SBM03\#3, GLARE\#3001} We adopted the measured
broadband fluxes and spectroscopic redshifts for these objects from
Tables 1 and 2 in~\citet{eyl05}.  Object SBM03\#1 is the same as object
\#1ab in~\citet{yan05}.  Comparing the derived fluxes in these two papers,
we find disagreement at the $\approx1\sigma$ level in the ACS $i_{775}$,
$z_{850}$, and NICMOS $H_{160}$ bands.  To be conservative, we therefore
impose a minimum uncertainty of 0.15 mag in all bands for these objects.
When fitting for the photometric redshifts we include the $i_{775}$
fluxes; otherwise we exclude this datum following~\citet{eyl05}.
Figure~\ref{fig:many_objs} shows that we obtain an excellent fit for
SBM03\#1 and satisfactory fits for SBM03\#3 and GLARE\#3001, with the
chief difficulty in the latter objects being the anomalous fluxes in $K_S$
as noted by~\citet{eyl05}.

For objects SBM03\#1~\citep{dic04,sta04a,sta03} and 
SBM03\#3~\citep{bun03,sta03} \spoc~deduces stellar masses of $\approx10^{10}
\msun$ and mean ages that include the range 100--300 Myr irrespective
of the assumed SFH, consistent with the findings of~\citet{eyl05}.
The minimum age for each model is older than the ages for the other
objects in our sample because the small measured uncertainties on the
3.6$\mu$m fluxes for SBM03\#1 and \#3 create the strongest case for a
pronounced Balmer break.  In our simulations, such objects are roughly
130--200 Myr old and are forming stars at a healthy 30--60 $\smyr$
(both $1\sigma$).  The resulting intrinsically blue colours cause \spoc~to 
select moderate dust extinctions $\Av$ of 0.4--0.9 and 0--0.1 for \#1 and 
\#3, respectively, in order to match their UV continuum slopes.  
While these SFRs are within the full range inferred via the one-parameter models, 
they are generally more active and younger than inferred via the decaying and
constant models.  The discrepancy owes primarily to the fact that the constant 
and decaying models permit higher SFRs  at early times ($z\ga10$) than occur 
in our simulations.  Because the rising model excludes such early episodes, 
it yields constraints that are more similar to the results from the jvzw 
models.

The photometric redshifts for SBM03\#1 and \#3 are 1--3$\sigma$ below the
spectroscopic values for both simulation and one-parameter models, suggesting that
unknown systematic uncertainties such as uncertainty owing to inaccurate
filter profiles should be folded into the formal uncertainties; in future
work an enforced minimum redshift uncertainty of $\delta z/(1+z)\geq 0.02$
seems reasonable.

Object GLARE\#3001 is a particularly interesting case because
it represents the worst fit for all our models.  This object
shows a relatively flat SED with little evidence for a Balmer
break (Figure~\ref{fig:many_objs}).  The photometric redshifts are
systematically high although they are accurate at the $1\sigma$ level.
\spoc\ finds that it is perhaps an order of magnitude less massive than
SBM03\#1 and \#3.  In our simulations, the analogues to GLARE\#3001
are of roughly the same age as the SBM objects while their SFRs
are roughly one tenth as large, leading to similarly strong Balmer
breaks and dust extinctions.  Comparing with the one-parameter model SFHs,
Table~\ref{table:many_objs} shows that, for this object, constant and
rising models yield masses, SFRs, and ages that are marginally consistent
with the results from our simulations (although they are skewed to higher
SFR) while decaying models yield extremely young, burst-like fits whose
parameters are entirely inconsistent with the jvzw results (this is 
especially clear in Figure~\ref{fig:degAll}).

The chief difficulty in fitting GLARE\#3001 is the anomalously high
flux measured in $K_S$, as noted by \citet{eyl05}.  If real, this
would suggest that this is a low-mass objects undergoing a burst.
Table~\ref{table:many_objs} shows that the $\chi_\nu^2$ per degree of
freedom is lower for the one-parameter models than for the simulated models,
because the one-parameter models have the freedom to produce a higher $K_S$
flux through a higher SFR and $\Av$ together with a lower stellar mass;
galaxies with these combinations of high SFR ($>10\smyr$) and low stellar
mass ($<10^9 \msun$) simply do not occur in our simulations.  It would
be preliminary to stake the final interpretation of this galaxy on one
broadband measurement (particularly one in $K_S$), especially since
even the one-parameter model fits are not particularly good ($\chi_\nu^2\ga 3$).
But this object does illustrate how \spoc\ can pick out galaxies that
may provide the most stringent tests of galaxy formation models.

\parname{Y05\#5abc,\#7ab} We adopted the \emph{Hubble}/ACS+NICMOS
and \emph{Spitzer}/IRAC fluxes for objects \#5abc and \#7ab from the
$z\approx6$ sample in~\citet{yan05} and imposed a minimum uncertainty
of 0.15 mag in all bands as before. \spoc~failed to find an acceptable
fit for object \#5abc, for either simulated or one-parameter model SFHs. This
object is evidently a blend of at least three components located
at different redshifts and should therefore be fit with multiple
components~\citep{yan05}; however, this is beyond the scope of the
present work.  We therefore do not show the results from \#5abc.

By contrast, we obtain excellent fits for object \#7ab.  This is
somewhat surprising given that \#7ab is clearly a mixture of two
components (Figure 1 in~\citet{yan05}), and probably owes largely
to the fact that it does not show anomalous single-band fluxes in
the way that HCM~6A, SBM03\#3, and GLARE\#3001 do.  The photometric
redshift is constrained to $6.2\pm0.2$, $1.5\sigma$ higher than the
photometric redshift determined by~\citet{yan05} (although they do not
quote an uncertainty).  Stellar mass is relatively poorly constrained
owing to the relatively low signal-to-noise in the observed 3.6$\mu$m
band.  As Figure~\ref{fig:degAll} shows, when we apply \spoc~with our
simulated models, the poorly-constrained stellar mass leads directly to
a poorly-constrained SFR.  Comparing with the one-parameter models, we find that
these span an even larger (overlapping) space.  This object illustrates
the importance of high-quality near-infrared data in order to understand
the physical properties of high-redshift galaxies.

\citet{yan05} applied exponentially-decaying models to this object
and found a burst-like best-fit solution with a stellar mass of
$\log(M_*/\msun) = 9.5$ (when converted to a Chabrier IMF), little current
star formation ($\sim0.001 \smyr$), no dust extinction, and an age of
50--100 Myr.  The stellar mass is fully consistent with our jvzw results.
However, we find disagreement in the inferred dust extinction, SFR, and
age because in the simulations galaxies of this stellar mass and redshift
are invariably older and are still forming stars.  In particular, we
expect $\Av = 0.34\pm0.24$, $\sfr = 9\pm5 \smyr$, and age = $175\pm30$~Myr
(1 $\sigma$ uncertainty).  Once again, the absence of burst-like models
in our simulations constitutes an effective prior that excludes such
solutions.  When we apply our own exponentially-decaying models to this
object we find that the probability density possesses a sharp peak at low
ages ($<100$ Myr) and a broad but slightly lower plateau at older ages.
So while we formally obtain the lowest $\chi^2$ for burst-like models,
folding in our simulation prior leads to an older age being preferred.
This object is therefore a classic case where simply taking the
lowest $\chi^2$ among all SFHs results in a substantially different
interpretation than that obtained using physically-motivated priors.
It will be interesting to see with improved obervations whether or not
the simulation prediction ends up being correct.

The sixth column of Table~\ref{table:many_objs} combines $\mstar$
and $\sfr$ with the age of the universe $t_H$ at each galaxy's redshift 
into an estimate of the ratio of the present- to past-averaged SFR, 
$\sfr/(\mstar/t_H)=\thtsfr$.  A value of $\thtsfr$ less than unity 
indicates a declining SFH whereas a value greater than unity indicates a
rising SFH.\footnote{Note that, because $\thtsfr$ does not involve the inferred
age of a galaxy's stars, this exercise is not circular; e.g., results from
fitting exponentially decaying models do not necessarily require that 
$\thtsfr < 1$.} \citet{eyl05,eyl06} have previously used this quantity to 
argue that galaxies that exhibit Balmer breaks at $z\sim6$ must be
experiencing a declining SFH.  In contrast, Table~\ref{table:many_objs}
indicates that, while this possibility is permitted by the decaying models 
at the $2\sigma$ level for all but one of the galaxies that we consider, it 
is by no means required\footnote{  
Decaying models permit $\thtsfr < 1$ while other models do not because 
decaying models can yield blue colors along with a lower specific SFR
than the other models.}.
For our physically-motivated models it is not permitted owing to the 
generally rising SFHs in our simulations, but in fact even for the decaying 
models it is not the favored solution when the full range of possibilities is 
taken into account.  Hence we do not find that the presence of a Balmer break 
requires a declining SFH.  The disagreement 
between our results and those of~\citet{eyl05} owes partly to the slightly 
different model sets used in the two analyses, and partly to our employing 
the full probability densities rather than restricting our attention to the 
model with the lowest $\chi^2$.  Most importantly, however, this 
disagreement emphasizes the fact that the form of the correct SFH is not 
constrained by the current measurements and hence the conclusions are highly
model-dependent.

In these fits, we have not considered the possibility of any AGN
contribution to the SEDs.  From rest-UV photometry alone it is
difficult to rule out this possibility.  Fitting power laws of the
form $f_\nu \propto \nu^\alpha$ to their rest-frame UV continua, we
obtain slopes in the range $\alpha \in [-0.3, -1.1]$ with a weighted
mean of -0.61.  These are not significantly different from the slopes
inferred by~\citet{fan01} from 39 quasars at $3.66 \leq z \leq 4.77$,
with a mean and standard deviation of -0.79 and 0.34, respectively.
However, with \emph{Spitzer}/IRAC photometry the Balmer breaks in A2218
KESR, SBM03\#1, SBM03\#3, and Y05\#7ab are clearly evident.  Furthermore,
\citet{sta04a} and~\citet{bun03} argue that AGN can be ruled out for three
of the objects (SBM03\#1, SBM03\#3, and GLARE\#3001) based on upper limits
to the flux of the \ion{N}{V} doublet at 1240\AA, the lack of any X-ray
detection, and the relatively small velocity width of the observed \lya
lines ($v_{\mbox{\tiny{FWHM}}}<500\kms$). These arguments also apply to
A370~HCM~6A.  Additionally, CSE05 note that the latter object is similar
to high-redshift \lya emitters, for which the AGN fraction is constrained
to be less than 5\%~\citep[e.g.,][]{gaw06,ouc05,daw04,wan04,san04b}.
This suggests that AGN do not dominate their rest-frame optical spectra,
and emphasizes the usefulness of \emph{Spitzer}/IRAC measurements for
constraining the properties of reionization-epoch objects.  Hence it
seems reasonable to model all of these objects with stellar population
synthesis models alone, as we have done.

In summary, we are usually but not always able to find galaxies within our
simulations whose photometry suggests that they could be the theoretical
counterpart to observed reionization-epoch galaxies.  Using these
models as inputs to our SED-fitting engine, \spoc, we obtain physical
constraints on the properties of the galaxies that are, in five out of
six cases, consistent with the results from one-parameter model SFHs.
In the remaining case, the observed SED leads the one-parameter models to
prefer burst-like scenarios that do not occur in our simulations.  In all
cases the tightness of constraints from the simulated models is equal to
or better than that from one-parameter model SFHs.  Albeit small, this
sample of systems illustrates how using \spoc\ to compare among SFHs can
provide insights into the physics of early galaxy formation, and identify
unusual galaxies that provide the tightest constraints on such models.

\section{Conclusions} \label{sec:summary}

In this paper we present a Bayesian SED-fitting engine called \spoc,
which provides constraints on the physical properties of galaxies from
photometric data.  \spoc\ takes as input a galaxy's photometry and a set
of model galaxy spectra, obtained either by assuming an analytic form
for the galaxy's star formation history or from numerical simulations of
galaxy formation, and outputs probability distributions for the physical
properties of individual galaxies based on those model priors.  Here we
compare and explore implications for different models, with an eye towards
better constraining the physical properties of high-redshift galaxies.

Because \spoc\ is intended to test whether predicted SFHs match those
inferred from data, it provides constraints on models of galaxy formation
that complement comparisons to bulk properties such as galaxy luminosity
functions or clustering.  A successful model must not only reproduce
observed bulk statistics, but must also reproduce the colors of individual
galaxies across all observed bands.  Galaxies with photometry that cannot
be well fit in a given model provide insights into model failings, and
ultimately into the physics that drives galaxy formation.  Using \spoc\
it is possible to quantitatively determine how well a particular model
matches an individual galaxy.  Such a methodology is particularly useful
at epochs when only a small number of galaxies are detectable, such as
for the earliest galaxies known.

We show that \spoc\ accurately recovers the input physical
parameters of model galaxies when fitted with SFH derived from the
simulations themselves, with typical systematic errors less than the
formal $1\sigma$ fitting errors.  When a simulated galaxy is fit with
one-parameter star formation histories or the incorrect extinction model,
then larger deviations can occur.  Since the true SFH and extinction law
are unknown, these deviations can be regarded as systematic errors on
the determination of physical parameters.  In most cases, such systematic
errors are less than 50\% in stellar mass, star formation rate, and $A_V$,
and $\la 2\%$ in redshift.

We then apply \spoc\ to six galaxies at $z\ga 5.5$ that have published
photometry spanning the 4000\AA\ break (which is necessary in
order to obtain meaningful constraints on physical properties).  We begin
with a more in-depth study of Abell 2218 KESR ($z\approx 6.7$), since
at present it is probably the best-studied reionization-epoch system,
and then apply \spoc\ to five more galaxies in order to investigate a
wider variety of galaxy properties.  Our main conclusions are summarized as
follows:

\begin{itemize}

\item The physical parameters derived for Abell 2218 KESR using our
library of simulated galaxies are in line with previous determinations,
and those employing various one-parameter SFHs.  The formal uncertainties
are smaller when using simulated galaxies, because the simulations
intrinsically produce a relatively narrow range of SFHs for galaxies at
these epochs.  The existence of an object with the properties of 
Abell 2218 KESR at these epochs does not pose a significant challenge
to current models of galaxy formation.

\item All inferred physical parameters for Abell 2218 KESR except
metallicity are remarkably insensitive to the explored choices
for superwind feedback model, dust extinction law, and cosmology.
The fundamental reason for this is that regardless of such choices,
simulations produce a similar {\it form} for the SFH of early galaxies.
Disappointingly, this means that photometry alone cannot constrain such
modeling uncertainties.  On the other hand, this means that simulation
predictions of physical parameters using \spoc\ are quite robust.
Discrimination between, say, superwind feedback models can be obtained
through comparisons with bulk properties such as luminosity functions
\citep[as shown in][]{dav06a}.

\item Exploring a set of six $z\ga 5.5$ galaxies, we find that for five
of the systems the simulated galaxies and one-parameter models produce
overlapping probability contours and best-fit physical parameters that are
formally consistent, and the minimum reduced $\chi^2$ is similar between
all models.  This shows that simulations are usually but not always able
to produce galaxies with properties similar to observed $z\sim 6$ systems.

\item The spectra of simulated galaxies always show a significant
Balmer break, despite the fact that their SFHs are best characterized
as constantly-rising (not decaying).  The median stellar age is $\approx
120-250$~Myr ($2\sigma$) for all six objects, spanning a range in stellar
masses from $\sim 5\times 10^8M_\odot \la M_*\la 2\times 10^{10}M_\odot$.
Hence the existence of somewhat older stellar populations in these
early systems is consistent with simulation predictions.

\item The object GLARE\#3001 is an outlier that is poorly fit by our
simulations, though in fairness the fits with one-parameter models are
not optimal either.  The best fits tend to favor models with younger
stellar ages, lower masses, and higher SFRs than any simulated galaxies,
perhaps indicating that this is a galaxy undergoing a burst.
However, as discussed in the text, these fits are mainly driven by
a high $K_s$-band flux.  If that data point were confirmed it would
suggest that the simulations may be missing some physical process(es)
that governs a small fraction of galaxies at this epoch.

\end{itemize}

Our hydrodynamic simulations of galaxy formation make some clear
predictions for the star formation histories of galaxies at these early
epochs.  In particular, they predict that the stellar mass, star formation
rate, and metallicity are all tightly correlated~\citep{fin06,dav06a}.
Indeed such trends are generically seen in most hydrodynamic simulations
of galaxy formation, though not necessarily in semi-analytic models.
Our preliminary comparisons to $z\sim 6$ galaxies show that, while the
simulations are not necessarily statistically favored over other classes
of SFHs, the available data at least do not rule out these simulation
predictions.  Together with the fact that such simulations can reproduce
key bulk properties of early galaxies~\citep{fin06,dav06a}, this suggests
that models of early galaxy formation are able to reproduce a range
of observed properties of galaxies at these epochs.

In this work we have performed the simplest possible spectral
synthesis calculations, primarily because our focus has been a
comparison between different SFH models.  In the future we plan to
investigate different population synthesis models~\citep[e.g.,][]{mar06},
IMFs~\citep[e.g.][]{far06}, and nonstellar contributions to the observed
fluxes such as emission lines from HII regions.  Eventually we will apply
\spoc\ to a larger sample of galaxies, which will enable a number of
interesting investigations.  First, the bulk statistics (e.g. stellar
mass or star formation rate functions) derived using \spoc\ can be
compared against that directly produced in simulations.  While this is
partly a circular comparison because simulated galaxies are being used
to infer the physical properties, in practice there is no guarantee that
consistency will be achieved because the simulated SFHs are quite generic;
hence this should provide a stringent test of models of galaxy formation.
Second, \spoc\ can be used to identify populations of galaxies that
deviate dramatically from simulation predictions (such as GLARE\#3001),
in order to isolate which physical processes may be missing in models.
In principle this could quantify the contributions from nonstellar
emission-dominated sources such as AGN and extremely dusty objects.
Finally, the redshift evolution of the galaxy population provides a
strong test of models, particularly investigating the dramatic change
in the nature of massive galaxies that appears to occur at $z\sim 1-2$
\citep{pap05}.

\spoc\ is a general-purpose code, in principle able to utilize any
type of model that produces detailed galaxy SFHs, be it a hydrodynamic,
semi-analytic, or analytic model.  In principle there is nothing that
restricts \spoc\ to high redshift use, but in practice other physical
phenomena such as AGN contamination may become more important at lower
redshifts.  \spoc\ could also provide a useful tool to identify galaxy
classes that are not readily reproduced in current models.  We are
currently planning to make \spoc~and our latest library of simulated
galaxies publicly available upon publication of this paper.  We hope
that \spoc\ will be a useful and flexible tool for conducting detailed
comparisons between simulations and observations, as are critical for
advancing our understanding of galaxy formation.

\section*{Acknowledgements}

We thank C.\ Papovich for a great deal of encouragement and an
inexhaustible stream of useful suggestions.  We also thank D.\
Zaritsky, E.\ Egami, R.\ Thompson, E.\ Bell, and A.\ Bunker for helpful
conversations, and V.\ Springel and L.\ Hernquist for providing us with
Gadget-2.  Finally, we thank the referee, F.\ Governato, for comments
that improved the paper.  The simulations were run in part on the 
Xeon Linux Supercluster
at the National Center for Supercomputing Applications, along with a
100-processor cluster here at Steward Observatory.  KMF acknowledges
support from a National Science Foundation Graduate Research Fellowship.
Support for this work was also provided by NASA through grant number
HST-AR-10647 from the Space Telescope Science Institute, which is operated
by the Association of Universities for Research in Astronomy, Inc., under
NASA contract NAS-26555.  Support for this work, part of the Spitzer
Space Telescope Theoretical Research Program, was further provided
by NASA through a contract issued by the Jet Propulsion Laboratory,
California Institute of Technology under a contract with NASA.

\onecolumn

\begin{deluxetable}{ccccccc}
\footnotesize
\tablecaption{Derived Parameters for Abell 2218 KESR}
\tablewidth{0pt}
\tablehead{
\colhead{Model\tablenotemark{a}} &
\colhead{$\Av$\tablenotemark{b}} &
\colhead{$Z_*$\tablenotemark{c}} &
\colhead{$Age$ (Myr)} &
\colhead{$\sfr$\tablenotemark{d} ($\smyr$)} &
\colhead{$\log(M_*/M_\odot)$\tablenotemark{d}} &
\colhead{$z$}
}
\startdata
nw  & $0.21 \pm 0.14$ & $0.0041 \pm 0.0007$ & $147 \pm 30$ & $2.2 \pm 0.6$ & $8.71 \pm 0.13$ & $6.71 \pm 0.06$\\
cw  & $0.14 \pm 0.12$ & $0.0035 \pm 0.0005$ & $160 \pm 33$ & $2.0 \pm 0.6$ & $8.73 \pm 0.14$ & $6.71 \pm 0.06$\\
vzw & $0.17 \pm 0.12$ & $0.0018 \pm 0.0002$ & $175 \pm 25$ & $2.2 \pm 0.6$ & $8.78 \pm 0.12$ & $6.71 \pm 0.06$\\
jvzw& $0.23 \pm 0.13$ & $0.0020 \pm 0.0002$ & $160 \pm 12$ & $2.4 \pm 0.7$ & $8.74 \pm 0.12$ & $6.71 \pm 0.06$\\
\citet{ega05} &       &                     & 50--450      & 0.1--3.3      & 8.5--8.8        & 6.6--6.8\\
\citet{sch05} &$\la 0.2$ &             &  3--400      & 1.5--1.9      & 7.7--9.0        & 6.0--7.2\\
\enddata
\tablenotetext{a}{nw, cw, vzw denote no, constant and momentum-driven winds, respectively; jvzw uses 3rd-year WMAP cosmology.}
\tablenotetext{b}{Assumes the~\citet{cal00} dust prescription and $R_V = 4.05$.}
\tablenotetext{c}{metal mass fraction in stars}
\tablenotetext{d}{Assumes a Chabrier IMF.  To convert from a Chabrier IMF to a
Salpeter IMF from 0.1--100 $\msun$, multiply by 1.5.}
\label{table:compareWind}
\end{deluxetable}

\begin{deluxetable}{lccccccc}
\footnotesize
\tablecaption{Mean and $2\sigma$ Parameter Ranges for $z\geq 5.5$ Galaxies}
\tablewidth{0pt}
\tablehead{
\colhead{Model\tablenotemark{a}} &
\colhead{$\Av$\tablenotemark{b}} &
\colhead{Age/Myr} &
\colhead{$\sfr$\tablenotemark{c} ($\smyr$)} &
\colhead{$\log(M_*/M_\odot)$\tablenotemark{c}} &
\colhead{$\thtsfr$\tablenotemark{d}} &
\colhead{$z$\tablenotemark{e}} &
\colhead{$\chi^2_{\nu}$\tablenotemark{f}}
}
\startdata
A2218 KESR  \\
jvzw      & 0.23 (0--0.53)  & 160 (132--181) & 2.4 (1.2--3.7)   & 8.7 (8.5--9.0) & 3.5 (2.5--4.2)  & $ 6.71\pm0.06$      &0.57 \\
constant  & 0.42 (0--1.0)   & 107 (7--350)   & 6.7 (2--25)      & 8.7 (8.1--9.1) & 17.2 (1.5--86.1) & $ 6.68\pm0.08$      &0.62 \\
decaying  & 0.34 (0--1.00)  & 104 (8--370)   & 6.9 (1.0--25.2)  & 8.7 (8.1--9.1) & 19.2 (0.5--86.0) & $ 6.68\pm0.07$      &0.55 \\
rising    & 0.33 (0--0.87)  & 129 (5--268)   & 3.5 (1.5--12.5)  & 8.7 (8.2--9.1) & 6.7 (2.2--56.3) & $ 6.69\pm0.07$      &0.60 \\
\cline{1-8}
A370 HCM\#6A \\
jvzw      & 0.49 (0--1.08)    & 191 (122--250) & 7.8 (2.7--18.2)   & 9.3 (8.9--9.6)  & 2.9 (2.3--4.4)   &$ 6.56/5.87\pm0.35$ &1.7 \\
constant   & 0.76 (0.13--1.24) & 182 (24--402)  & 14.0 (4.1--28.1)  & 9.6 (9.0--9.9)  & 3.9 (1.1--17.3) &$ 6.56/6.18\pm0.41$ &1.5 \\
decaying   & 0.55 (0--1.37)    & 217 (9--502)   & 16.4 (0.6--83.5)  & 9.5 (8.7--9.9)  & 9.2 (0.4--86.1) &$ 6.56/6.15\pm0.41$ &1.5 \\
rising     & 0.82 (0.13--1.37) & 138 (8--282)   & 16.1 (4.2--40.1)  & 9.4 (8.8--9.8)  & 6.1 (2.2--43.4)     &$ 6.56/6.15\pm0.41$ &1.6 \\
\cline{1-8}
SBM03\#1 \\
jvzw      & 0.67 (0.36--0.93) & 175 (123--231) & 42.1 (30.9--62.1)  & 10.1 (9.9--10.2)  & 3.7 (2.4--4.7) &$ 5.83/5.64\pm0.08$ &1.7 \\
constant  & 0.45 (0.25--0.62) & 388 (235--498) & 24.5 (18.0--29.3)  & 10.2 (10.1--10.4) & 1.4 (1.0--2.1) &$ 5.83/5.68\pm0.08$ &1.2 \\
decaying  & 0.29 (0--0.75)    & 357 (89--601)  & 20.7 (0.63--99.6)  & 10.3 (10.0--10.5) & 1.3 (0.1--10.1) &$ 5.83/5.68\pm0.08$ &0.9 \\
rising    & 0.78 (0.50--1.12) & 212 (51--332)  & 51.7 (29.5--102.0) & 10.2 (10.0--10.4) & 3.4 (2.0--9.0)   &$ 5.83/5.65\pm0.08$ &1.6 \\
\cline{1-8}
SBM03\#3 \\
jvzw      & 0.44 (0.09--0.69) & 170 (124--231) & 46.1 (24.9--75.3)  & 10.1 (9.9--10.3)  & 3.9 (2.4--4.8) &$ 5.78/5.68\pm0.07$ &2.2 \\
constant  & 0.18 (0--0.37)    & 385 (225--498) & 25.0 (18.0--29.3)  & 10.3 (10.0--10.4) & 1.4 (1.0--2.3) &$ 5.78/5.74\pm0.07$ &1.7 \\
decaying  & 0.25 (0--0.87)    & 301 (30--601)  & 41.9 (8.3--251.0)  & 10.3 (9.9--10.6)  & 3.0 (0.2--20.4) &$ 5.78/5.71\pm0.08$ &1.5 \\
rising    & 0.58 (0.25--1.00) & 188 (25--332)  & 64.9 (28.4--159.2) & 10.2 (9.9--10.4)  & 4.4 (2.1--13.7) &$ 5.78/5.69\pm0.08$ &2.0 \\
\cline{1-8}
GLARE\#3001 \\
jvzw      & 0.19 (0--0.46)    & 183 (123--253) & 5.2 (3.4--8.7)     & 9.2 (9.0--9.3)   & 3.7 (2.4--5.9) &$ 5.79/5.86\pm0.07$ &8.7 \\
constant  & 0.47 (0.12--0.62) & 31 (6--126)    & 18.9 (6.2--29.3)   & 8.8 (8.4--9.2)   & 37.3 (4.5--101.2) &$ 5.79/5.81\pm0.07$   &3.5 \\
decaying  & 0.59 (0.13--0.87) & 15 (6--85)     & 30.3 (7.4--63.3)   & 8.7 (8.4--9.2)   & 63.5 (5.6--101.2) &$ 5.79/5.80\pm0.08$   &3.3 \\
rising    & 0.53 (0--0.87)    & 42 (5--200)    & 17.6 (5.1--40.1)   & 8.8 (8.4--9.3)   & 38.1 (3.4--101.2) &$ 5.79/5.81\pm0.08$   &3.0 \\
\cline{1-8}
Y05\#7ab \\
jvzw        & 0.34 (0--0.94) & 175 (126--227) & 9.2 (3.7--26.7)   & 9.4 (9.0--9.8)    & 3.4 (2.6--4.8) &$ 6.25\pm0.16 $ &0.56 \\
constant    & 0.56 (0--1.00) & 210 (27--448)  & 16.2 (4.7--28.7)  & 9.7 (8.8--10.3)   & 4.3 (1.2--19.0) &$ 6.17\pm0.17 $ &0.18 \\
decaying    & 0.48 (0--1.24) & 212 (10--532)  & 22.7 (0.8--105.3) & 9.7 (8.5--10.4)   & 12.4 (0.5--92.8) &$ 6.15\pm0.18 $ &0.14 \\
rising      & 0.63 (0--1.37) & 140 (5--299)   & 20.3 (4.6--63.8)  & 9.5 (8.7--10.2)   & 7.5 (2.2--61.1) &$ 6.18\pm0.18 $ &0.35 \\
\enddata
\tablenotetext{a}{``jvzw" denotes the simulated galaxies; ``constant", ``decaying", 
and ``rising" denote representative one-parameter model SFH models.}
\tablenotetext{b}{Assumes the~\citet{cal00} dust prescription and $R_V =
4.05$.}
\tablenotetext{c}{Assumes a Chabrier IMF.  To convert from a Chabrier IMF to a
Salpeter IMF from 0.1--100 $\msun$, multiply by 1.5.}
\tablenotetext{d}{The ratio of the current SFR to the past-averaged SFR,
defined as $\thtsfr$}
\tablenotetext{e}{Where 2 numbers are given, the first is the spectroscopic
redshift and the second is the photometric redshift.}
\tablenotetext{f}{Minimum reduced $\chi^2$ for each combination of models and
observed object.  The number of degrees of freedom is given by the number of detections
used minus 3 for A2218 KESR and Y05\#7ab, and minus 2 for the others.}
\label{table:many_objs}
\end{deluxetable}
\twocolumn

\end{document}